\documentclass[aps,prb,nobibnotes,superscriptaddress,reprint]{revtex4-1}
\usepackage{bm}
\usepackage{graphicx}
\usepackage{amsmath}
\usepackage{amssymb}
\usepackage{dcolumn} 
\usepackage[colorlinks=true,citecolor=blue,linkcolor=blue,urlcolor=blue]{hyperref}
\usepackage{subcaption}
\usepackage{xspace}
\usepackage{paralist}

\newcommand{\fref}[1]{Fig.~\ref{#1}}
\newcommand{\tref}[1]{Table~\ref{#1}}
\newcommand{\eref}[1]{(\ref{#1})}
\newcommand{\sref}[1]{Section~\ref{#1}}
\newcommand{\ipfullname}[0]{Dihedral-angle-corrected Registry-dependent Interlayer Potential\xspace}
\newcommand{\ipshortname}[0]{DRIP\xspace}
\newcommand{\eqn}[1]{Eq.~\eqref{#1}}
\newcommand{\eqns}[1]{Eqs.~\eqref{#1}}

\newcommand{\mc}[1]{\textcolor{black}{#1}}
\newcommand{\mcr}[1]{\textcolor{black}{#1}}

\graphicspath{{figures/}{./}}

\begin{document}

\title{Dihedral-angle-corrected registry-dependent interlayer potential for multilayer graphene structures}

\author{Mingjian Wen}
\affiliation{Department of Aerospace Engineering and Mechanics, University of Minnesota, Minneapolis, MN 55455, USA}

\author{Stephen Carr}
\affiliation{Department of Physics, Harvard University, Cambridge, MA 02138, USA}

\author{Shiang Fang}
\affiliation{Department of Physics, Harvard University, Cambridge, MA 02138, USA}

\author{Efthimios Kaxiras}
\affiliation{Department of Physics, Harvard University, Cambridge, MA 02138, USA}
\affiliation{John A. Paulson School of Engineering and Applied Sciences, Harvard University, Cambridge, MA 02138, USA}

\author{Ellad B. Tadmor}
\email[Author to whom correspondence should be addressed: ]{tadmor@umn.edu}
\affiliation{Department of Aerospace Engineering and Mechanics, University of Minnesota, Minneapolis, MN 55455, USA}

\date{\today}

\begin{abstract}
The structural relaxation of multilayer graphene is essential in describing
the interesting electronic properties induced by intentional misalignment of
successive layers, including the recently reported superconductivity
in twisted bilayer graphene.   This is difficult to accomplish without an
accurate interatomic potential.  Here, we present
a new, registry-dependent Kolmogorov-Crespi type
interatomic potential to model interlayer interactions in
multilayer graphene structures.
It consists of two parts representing attractive interaction due to dispersion,
and repulsive interaction due to anisotropic overlap of electronic orbitals.
An important new feature is
a dihedral-angle-dependent term that is added to the repulsive part in order to
describe correctly several distinct stacking states that the original
Kolmogorov-Crespi potential cannot distinguish.
We refer to the new model as the \ipfullname (\ipshortname).
Computations for several test problems show that \ipshortname correctly reproduces the binding, sliding, and twisting energies and forces obtained from
\emph{ab initio} total-energy calculations based on
density functional theory.
We use the new potential to study the structural properties
of a twisted graphene bilayer and the exfoliation of graphene from graphite.
Our potential
is available through the OpenKIM interatomic potential repository at \url{https://openkim.org}.

\end{abstract}

\maketitle

\section{Introduction}
\label{sec:intro}

Since the discovery of graphene\cite{novoselov2004electric}, two-dimensional (2D) materials have been shown to possess remarkable electronic, mechanical, thermal, and optical properties with great potential for nanotechnology applications, such as semiconductors, ultrasensitive sensors, and medical devices\cite{neto2009electronic,hendry2010coherent,sevik2014assessment,lee2008measurement}.
Stacked 2D materials (or ``heterostructures'') have even more unusual and novel properties that their monolayer and 3D counterparts do not possess.\cite{geim2013van,novoselov20162d}
For example, the electronic band gap of a graphene bilayer can be tuned
by applying a variable external electric field, which allows great flexibility in the design and optimization of semiconductor devices such as p-n junctions and transistors.\cite{zhang2009direct}
A different manifestation of interesting behavior not found in the bulk is the
recently reported superconductivity in intentionally misaligned
(by a relative twist of $\sim 1.1^\circ$) graphene bilayers\cite{graphene_sc_2018}.
As a prototype of a stacked 2D material, multilayer graphene (``graphitic structure'' hereafter) exhibits strong $sp^2$ covalent bonds within layers and weak van der Waals (vdW) and orbital repulsion interactions between layers.
Although weak, it is the interlayer interaction that defines the function of nanodevices such as nanobearings, nanomotors and nanoresonators.\cite{kolmogorov2005registry}

To simulate the mechanical behavior of graphitic structures
it is necessary to model the interactions between the electrons and the ions, which
produce the forces governing atomic motion and deformation.
First-principles approaches that involve solving the Schr\"{o}dinger equation are most accurate, but due to hardware and algorithmic limitations,
this approach is typically limited to studying small molecular systems and crystalline materials characterized by compact unit cells with
an upper limit on the number of atoms in the range of $\sim 10^3$.
Empirical interatomic potentials are computationally far less costly than
first-principles methods and can therefore be used to compute static
and dynamic properties that are inaccessible to quantum calculations,
such as dynamical tribological properties of large-scale graphene
interfaces.\cite{wen2015interpolation, leven2014ilp, leven2016interlayer}

There have been many efforts to produce an interatomic potential that would
adequately describe the properties of graphitic structures, in particular the interactions between layers.
However, as we argue in detail in this paper, the existing potentials fall short
of capturing key elements of the graphitic structures of interest.
Therefore, there is a pressing need to construct an accurate interlayer potential that will
elucidate many of the important structural properties of these structures.

The paper is structured as follows.
In \sref{sec:need} we briefly review the nature of existing
interatomic potentials that might be applied to graphitic structures,
we explain their shortcomings, and
elaborate on the need for constructing a new potential.
In \sref{sec:model}, the functional form of the new model is
presented, together with
a description of the fitting process that determines the values of
all the parameters that appear in it.
In \sref{sec:test}, the predictions of the new model for several canonical
properties of interest are compared with other potentials and
results from \emph{ab initio} total-energy calculations based
on density functional theory (DFT).
Large-scale applications of the new model are discussed in \sref{sec:app}.
The paper is summarized in \sref{sec:sum}.

\section{Need for new graphitic potential
\label{sec:need}
}

\begin{figure*}
\begin{subfigure}{0.65\columnwidth}
\includegraphics[width=\columnwidth]{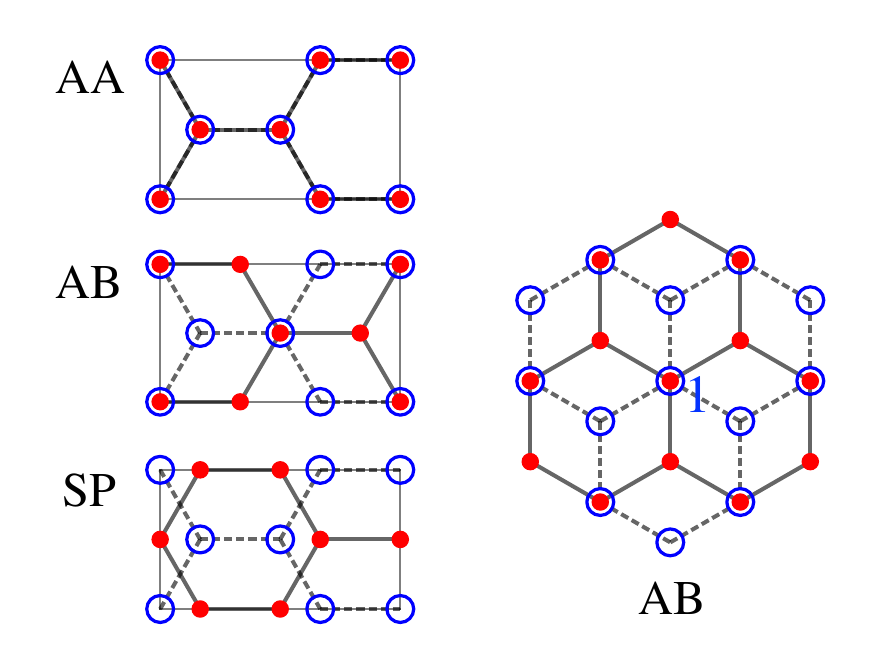}
\caption{}
\label{fig:high:sym:confs}
\end{subfigure}
\begin{subfigure}{0.65\columnwidth}
\includegraphics[width=\columnwidth]{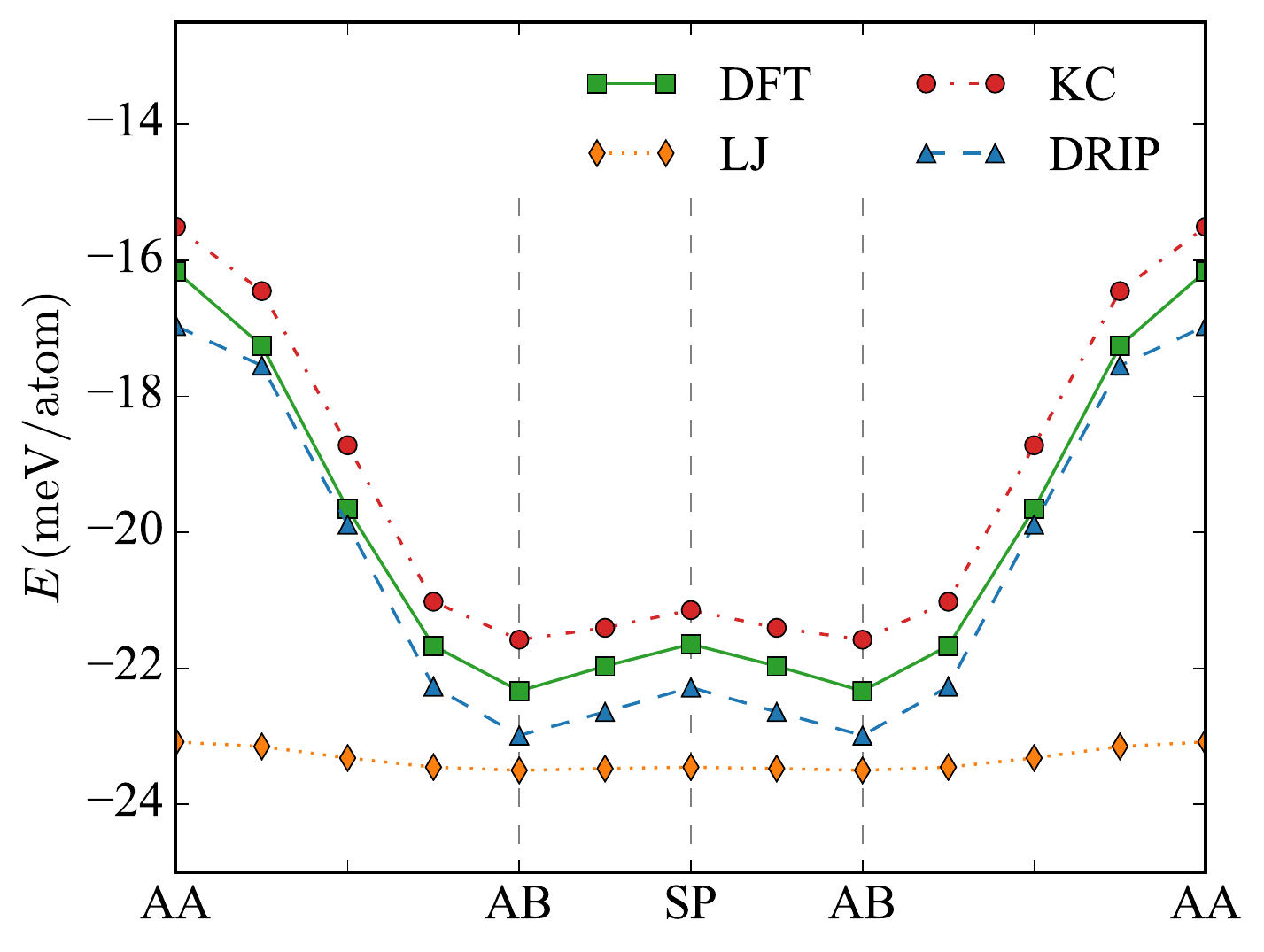}
\caption{}
\label{fig:sliding:energy}
\end{subfigure}
\begin{subfigure}{0.65\columnwidth}
\includegraphics[width=\columnwidth]{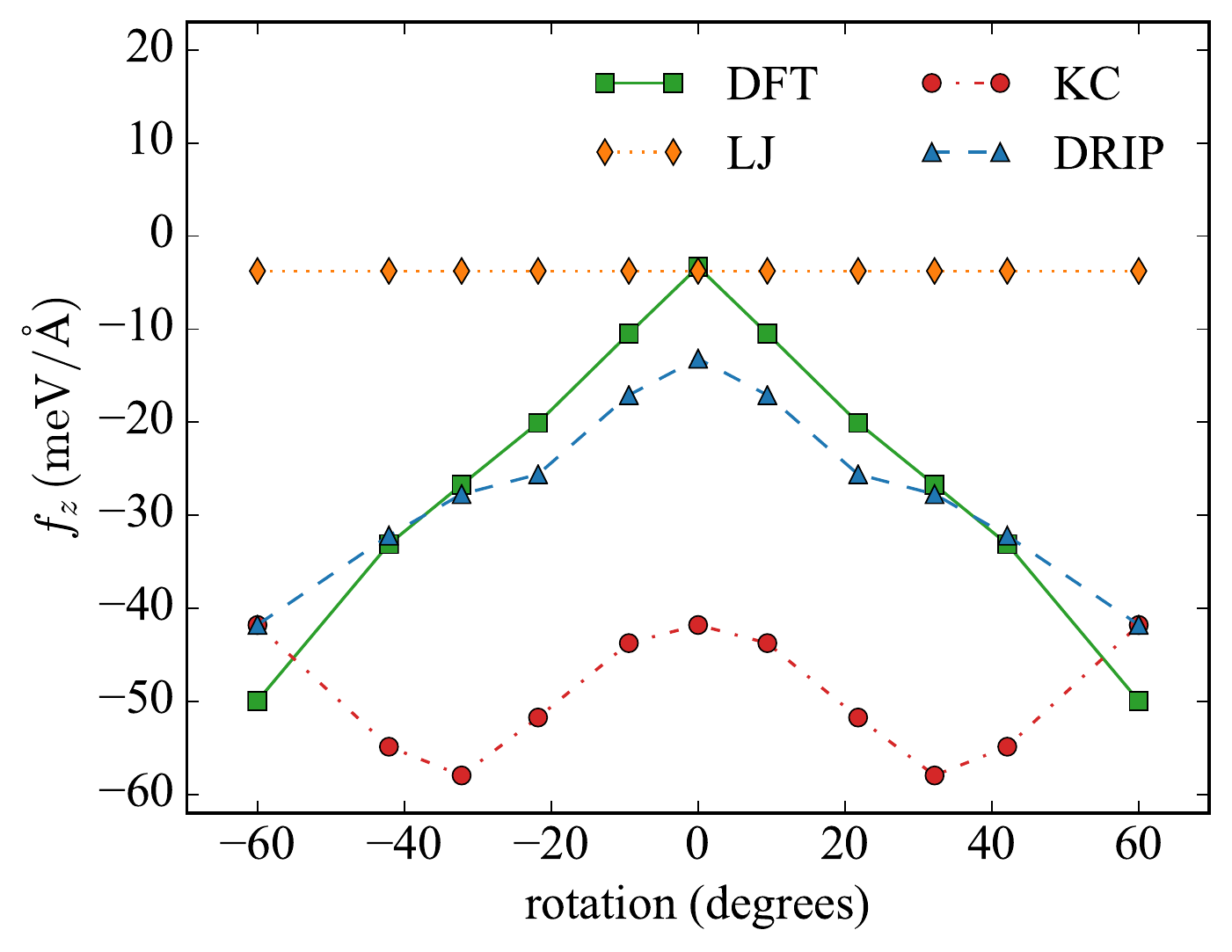}
\caption{}
\label{fig:rotation:force}
\end{subfigure}
\caption{Energy and force variations when sliding and twisting a graphene bilayer.
(a) Schematic representation of high-symmetry graphene bilayer configurations: AA, AB, and saddle point (SP) stacking.
(b) Energy variation of sliding one layer relative to the other along the armchair direction.
(c) Out-of-plane component of the force on the atom at the rotation center
(blue circle labeled 1 in the bottom layer in panel (a)).
Rotation by $0^\circ$ corresponds to AB stacking,
and rotation by $\pm60^\circ$ corresponds to AA stacking.
In both sliding and twisting, periodic boundary conditions are applied and the layer spacing is fixed at $3.4~\text{\AA}$. Details are provided in \sref{sec:test}.}
\end{figure*}

A large number of interatomic potentials have been developed to model the strong covalent bonds in carbon systems. Among these are bond-order potentials, such as the Tersoff\cite{tersoff1988empirical,tersoff1989modeling} and REBO\cite{brenner1990physical,brenner2002rebo} potentials, which allow for bond breaking and formation depending on the local atomic environments. Such models have been shown to be accurate for many problems and are widely used,
but are not suitable for layered 2D materials since
they do not include long-range weak interactions.
To address this, the AIREBO\cite{stuart2000airebo} potential (based on REBO) added a \mcr{6-12 form of the} Lennard-Jones (LJ) potential\cite{lennardjones1931} to model vdW interactions.
For graphitic structures, the LJ potential works well in describing the overall binding characteristics between graphene layers.
For example, the LJ parameterization used in AIREBO predicts an equilibrium layer spacing of $3.357~\text{\AA}$ and a $c$-axis elastic modulus of $37.78~\text{GPa}$ for graphite,
in good agreement with first-principles and experimental results.
The isotropic nature of LJ, that is,
the fact that it depends only on distance between atoms and not orientation,
makes it too smooth to distinguish energy variations for different relative alignments of layers.\cite{zhang2017energy}
\fref{fig:sliding:energy} shows the energy variation obtained by sliding one layer relative to the other along the armchair direction of a graphene bilayer.
The energy remains nearly constant with a maximal difference of $0.41~\text{meV/atom}$ between the AA and AB stackings,
a small fraction (6.6\%) of the DFT result (also shown in the figure).

The reason that the LJ potential fails to capture the energy variations
due to interlayer sliding
is that in addition to vdW, the interlayer interactions include
short-range Pauli repulsion between overlapping $\pi$ orbitals of adjacent layers.
These repulsive interactions are not well described by a simple pair potential like LJ.\cite{kolmogorov2005registry,leven2014ilp,leven2016interlayer} To account for this registry effect (relative alignment of layers), Kolmogorov and Crespi (KC) developed a registry-dependent interlayer potential for graphitic structures.\cite{kolmogorov2005registry}
In the KC potential, the dispersive (vdW) attraction between layers is described
using the same theoretically-motivated $r^{-6}$ term as in LJ,
and $\pi$ orbital overlap is modeled by a Morse\cite{morse1929diatomic} type exponential multiplied by a registry-dependent modifier that depends on the transverse distance between atom pairs.
\mc{The KC potential has been modified and reparameterized to better fit the energy
variations between different stacking states predicted by DFT-D (DFT with
dispersion corrections)\cite{lebedeva2011interlayer}.}
It has also been adapted for other 2D materials such as h-BN\cite{leven2014ilp}
and graphene/h-BN\cite{leven2016interlayer, maaravi2017interlayer} heterostructures.

\mc{The energy corrugation obtained by the KC potential is in agreement with
DFT as shown in \fref{fig:sliding:energy}.}
However, the forces obtained from the KC potential deviate significantly from the DFT results. This implies that equilibrium structures associated with
energy minima will differ as well.
To illustrate this point, consider a graphene bilayer where one layer is
rigidly rotated relative to the other.  \fref{fig:rotation:force} shows the force
in the $z$-direction (perpendicular to the layers) acting on the bottom atom on
the rotation axis (atom~1 in the bottom layer in \fref{fig:high:sym:confs}) as
a function of rotation angle.
The force predicted by the KC potential decreases and then increases from
AA ($\pm60^\circ$) to AB ($0^\circ$), whereas DFT predicts a monotonic increase from AA to AB\@.
In particular, the KC potential yields the same $z$-force for the AA and AB stackings\footnote{Note that the $x$ and $y$ components of the force are zero at AA and AB due to symmetry}, which indicates that the KC potential cannot distinguish the overlapping atoms at the rotation center in these states.
This is intrinsic to the KC potential. The force on the central atom in the AA and AB states is identical, regardless of the choice of KC parameters.
The modified KC potential\cite{lebedeva2011interlayer} has the same problem.
The LJ potential does even worse (\fref{fig:rotation:force}) predicting a constant
force on the central atom that is independent of the rotation angle.

In the present paper, a new registry-dependent interlayer potential for graphitic structures is developed that addresses the limitations of the KC potential described above.
A dihedral-angle-dependent term is introduced into the registry modifier of the repulsive part that makes it possible to distinguish forces in AA and AB states.
We refer to this potential as the \ipfullname (\ipshortname).
\ipshortname is validated by showing that it
correctly reproduces the DFT energy and forces for different sliding and rotated states as well as structural and elastic properties.
It is then applied to study structural relaxation in twisted graphene bilayers
and exfoliation of graphene from graphite; these representative example are
large-scale applications that cannot be studied using DFT.
The potential has been implemented as a \ipshortname Model Driver\cite{drip_driver} and the parameterization in this paper has been implemented as a Model\cite{drip_model} at OpenKIM\cite{tadmor2011kim,tadmor2013nsf}.
(See details in
Appendix~\ref{appendix:kim}.)

\section{Definition of new model}
\label{sec:model}

The \ipshortname functional form is
\begin{equation}\label{eq:tot:eng}
  \mathcal{V} = \frac{1}{2} \sum_{i \in \text{layer 1}} \sum_{j  \in \text{layer 2}} (\phi_{ij} + \phi_{ji}) ,
\end{equation}
where the pairwise interaction is based on the KC form with dihedral modifications:
\begin{widetext}
\begin{equation}
\phi_{ij} = f_\text{c}(x_r) \left[ e^{-\lambda(r_{ij} - z_0 )} \left[C+f(\rho_{ij})
+  g(\rho_{ij}, \{\alpha_{ij}^{(m)}\}) \right]
- A\left (\frac{z_0}{r_{ij}} \right)^6 \right], \; \;
m=1,2,3.
\label{eq:phi}
\end{equation}
\end{widetext}
The cutoff function $f_\text{c}(x)$ is same as that used in the ReaxFF potential\cite{duin2001reaxff} and the interlayer potential for h-BN\cite{leven2014ilp, leven2016interlayer}:
\begin{equation} \label{eq:cutoff}
f_\text{c}(x)=
  20x^7 - 70x^6 + 84x^5 - 35x^4 +1,
\end{equation}
for $0\leq x \leq 1$ and vanishes for $x>1$,
while it has zero first and second derivatives at $x=1$;
in the expressions where
this function appears its argument is always non-negative.
The variable $x_r$ in \eqn{eq:phi} is the scaled pair distance
$x_r = r_{ij}/r_\text{cut}$.
The use of $f_\text{c}(x_r)$ ensures that \ipshortname is smooth at the cutoff $r_\text{cut}$ (set to 12~\AA), a feature that the KC model does not possess.

\begin{figure}
\includegraphics[width=0.60\columnwidth]{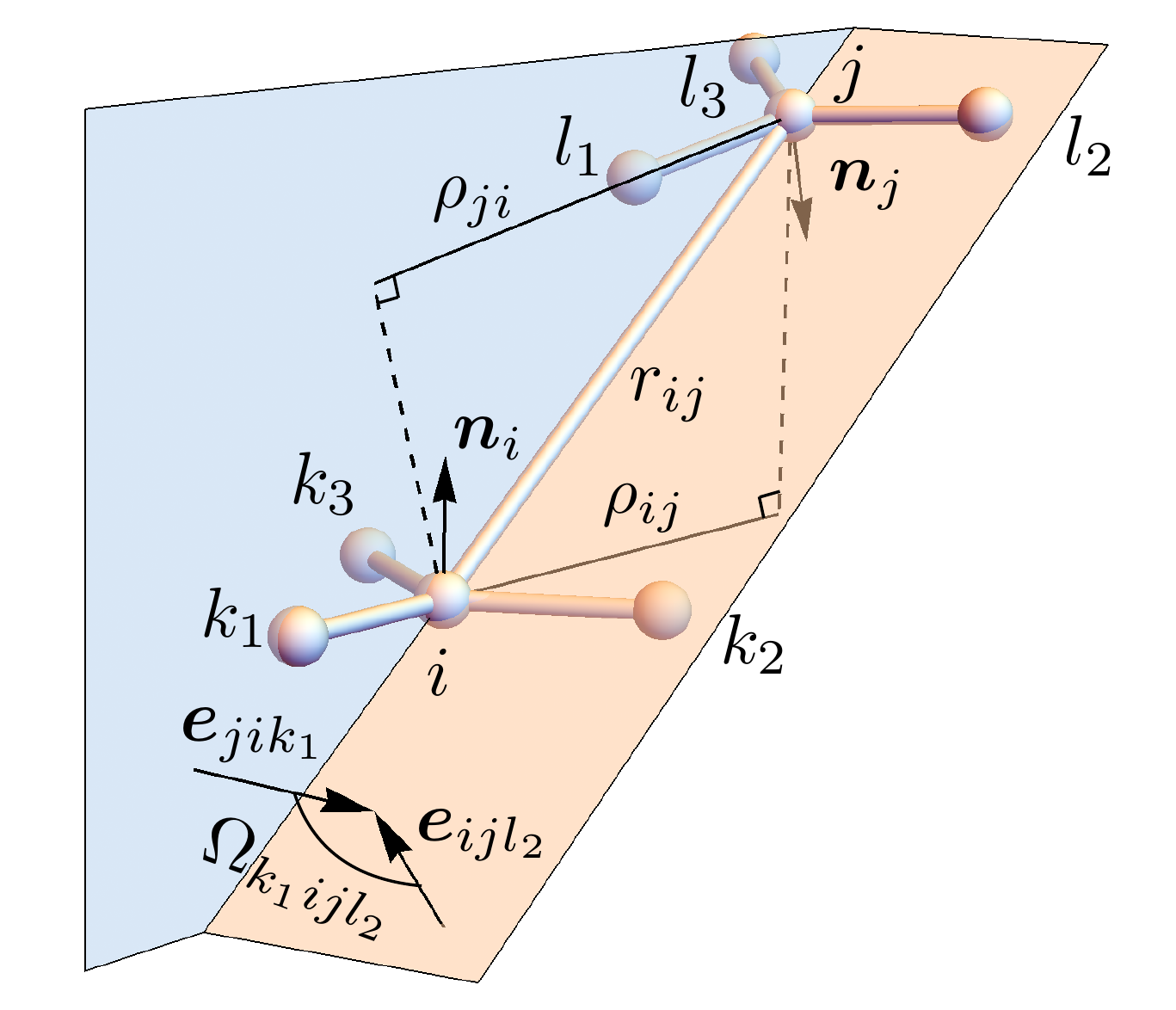}
\caption{Schematic representation of an atomic geometry that
defines the normal vectors $\bm n_i$
and $\bm n_j$ and the dihedral angle $\Omega_{k_1 ijl_2}$.}
\label{fig:normal:dihedral}
\end{figure}

The term with $r_{ij}^{-6}$ dependence in
\eqn{eq:phi} models attractive vdW interactions (as in LJ),
while the repulsive interactions due to orbital overlap are modeled by
the exponential term multiplied by a registry-dependent modifier.
The transverse distance function $f(\rho)$ has the same form as in KC:
\begin{equation}
f(\rho) = e^{-y^2} \left[ C_0 +  C_2 y^2 + C_4 y^4 \right],
\; \; y=\frac{\rho}{\delta}
\end{equation}
with its argument in \eqn{eq:phi} given by the expression
\begin{equation}
\rho_{ij}^2 = r_{ij}^2 - (\bm n_i \cdot \bm r_{ij})^2,
\end{equation}
in which $\bm r_{ij}$ is the vector connecting atoms $i$ and $j$, $r_{ij}$ is the corresponding pair distance, and $\bm n_i$ is the layer normal at atom $i$.
For example, as shown in \fref{fig:normal:dihedral}, $\bm n_i$ can be defined as the normal to the plane determined by the three nearest-neighbors of atom $i$: $k_1, k_2$ and $k_3$:
\begin{equation}
  \bm n_i =
\frac{\bm r_{k_1k_2}\times \bm r_{k_1k_3}}{\| \bm r_{k_1k_2}\times \bm r_{k_1k_3} \|}.
\end{equation}
Note that in general $\rho_{ij} \neq \rho_{ji}$ because the normals $\bm n_i$ and $\bm n_j$ depend on their local environments.

The dihedral angle function is given by
\begin{equation} \label{eq:g}
g(\rho, \{\alpha_{ij}^{(m)}\}) = B f_\text{c}(x_\rho) \sum_{m=1}^3 e^{-\eta \alpha_{ij}^{(m)}},
\end{equation}
where $\alpha_{ij}^{(m)}$ is the product of the three cosines of the dihedral angles formed by atom $i$ (in layer~1), its $m$th nearest-neighbor $k_m$, atom $j$ (in layer~2), and its three nearest-neighbors $l_1, l_2$ and $l_3$:
\begin{equation} \label{eq:alpha}
\alpha_{ij}^{(m)} = \cos\Omega_{k_mijl_1} \cos\Omega_{k_mijl_2} \cos\Omega_{k_mijl_3}
\end{equation}
\begin{equation} \label{eq:cos:dihedral}
\cos\Omega_{kijl} = \bm e_{jik} \cdot \bm e_{ijl}
\end{equation}
\begin{equation} \label{eq:normal:to:plane}
\bm e_{jik} = \frac{\bm r_{ik} \times \bm r_{ji}} {\|\bm r_{ik} \times \bm r_{ji}\|}, \qquad
\bm e_{ijl} = \frac{\bm r_{jl}  \times \bm r_{ij}} {\|\bm r_{jl}  \times \bm r_{ij}\|}.
\end{equation}
To understand the physical origin of the terms defined in
\eqns{eq:alpha}--\eqref{eq:normal:to:plane}, recall that
a dihedral angle $\Omega$ is the angle between two planes defined by four points that intersect at a line defined by two of them as shown in \fref{fig:normal:dihedral}.
Here, the intersection line is defined by atoms $i$ and $j$.
The two planes are then defined by atoms $(j, i, k_1)$ and $(i, j, l_2)$.
The normals to these planes are $\bm e_{jik_1}$ and $\bm e_{ijl_2}$, respectively,
defined in \eqn{eq:normal:to:plane}, with
the corresponding dihedral angle given by \eqn{eq:cos:dihedral}.
The dihedral product $\alpha_{ij}^{(m)}$ monotonically decreases when twisting a graphene bilayer from AB to AA stacking, and consequently can be utilized to construct a potential function that distinguishes AB and AA stacking and the intermediate stacking states.
The cutoff function $f_\text{c}(x_\rho)$ in
\eqn{eq:g} is the same as that in \eqn{eq:cutoff}, and
$x_\rho = \rho/\rho_\text{cut}$,
where we set $\rho_\text{cut} = 1.562~\text{\AA}$ to include only a few
of the computationally expensive 4-body dihedral angle interactions.
The potential has a total of
ten parameters, $C_0$, $C_2$, $C_4$, $C$, $\delta$, $\lambda$, $B$, $\eta$, $A$, and $z_0$, and two cutoffs $r_\text{cut}$ and $\rho_\text{cut}$.


To determine the values of all the parameters that appear in the
\ipshortname potential, we constructed
a training set of energies and forces for graphene bilayers
at different separation, sliding, and twisting states.
The training set is generated from DFT calculations using the Vienna Ab initio Simulation Package (VASP)\cite{vasp1,vasp2}.
The exchange-correlation energy of the electrons is treated within the generalized gradient approximated (GGA) functional of Perdew, Burke and Ernzerhof (PBE)\cite{pbe}.

Standard density functionals such as the local density approximation (LDA) and GGA accurately represent Pauli repulsion in interlayer interactions,
but fail to capture vdW forces that result from dynamical correlations between fluctuating charge distributions.\footnote{GGA predicts no binding at all at physically meaningful spacings for graphite. LDA gives the correct interlayer spacing for AB stacking, however, it underestimates the exfoliation energy by a factor of two and overestimates the compressibility.\cite{kolmogorov2005registry}}
To address this limitation,
various approximate corrections have been proposed including the D2 method\cite{grimme2006semiempirical}, the D3 method\cite{grimme2010consistent},
the Tkatchenko and Scheffler (TS) method\cite{tkatchenko2009accurate},
the TS method with iterative Hirshfeld partitioning (TSIHP) method\cite{bucko2013improved}, the many-body dispersion (MBD) method\cite{tkatchenko2012accurate},
and the dDsC dispersion correction method\cite{steinmann2011generalized}.
To select a correction for the \ipshortname training set, we used these dispersion
correction methods to compute a number of structural, energetic, and elastic properties.
\mc{The results are shown in \tref{tab:properties} along with experimental
values and more accurate adiabatic-connection fluctuation-dissipation theory
based random-phase-approximation (ACFDT-RPA) computations that have been shown
to provide a very accurate description of vdW interactions\cite{zhou2015van,lebegue2010cohesive}.
The conclusion from these comparisons is that
D2 and D3 provide inaccurate estimates for the layer spacing of AB graphene and graphite
($d_\text{AB}$ and $d_\text{graphite}$), and TS, TSIHP, and dDsC significantly
overestimate the graphite binding energy $E_\text{graphite}$.
MBD provides the best overall accuracy for all considered properties and is therefore
the vdW correction used in this work together with the PBE functional.}

\begin{table*}
\caption{Properties obtained from various DFT vdW corrections compared with
ACFDT-RPA and experimental results.
\mc{
Also included are results from various empirical potentials.
The properties include:
equilibrium layer spacings of bilayer graphene in AB stacking, $d_\text{AB}$,
bilayer graphene in AA stacking, $d_\text{AA}$,
and graphite, $d_\text{graphite}$;
optimal interlayer binding energies for bilayer graphene (binding energy at the
equilibrium spacing in AB stacking), $E_\text{AB}$,
and graphite, $E_\text{graphite}$;
energy differences between AA-stacked and AB-stacked bilayers, $\Delta E_\text{AA-AB}$,
and SP and AB stackings, $\Delta E_\text{SP-AB}$, at a layer spacing of $d=3.4~\text{\AA}$;
and the elastic modulus along the $c$-axis for graphite, $C_{33}$.
All properties are computed using the in-plane lattice constant $a=2.46~\text{\AA}$.
}
}
\label{tab:properties}
\begin{ruledtabular}
\begin{tabular}{ccccccccc}
     &$d_\text{AB}$  &$d_\text{AA}$  &$d_\text{graphite}$  &$E_\text{AB}$
     &$E_\text{graphite}$  &$\Delta E_\text{AA-AB}$  &$\Delta E_\text{SP-AB}$  &$C_{33}$\\
     & (\AA)         &(\AA)          &(\AA)                &(meV/atom)
     &(meV/atom)           &(meV/atom)            &(meV/atom)           &(GPa)  \\
\hline
PBE+D2     &3.248  &3.527  &3.218  &24.84  &55.20  &10.35  &1.16    &39.12  \\
PBE+D3     &3.527  &3.713  &3.483  &21.40  &47.09  &3.80  &0.42    &35.04  \\
PBE+TS     &3.357  &3.511  &3.329  &36.36  &82.33  &7.97  &1.01    &68.31  \\
PBE+TSIHP  &3.379  &3.529  &3.350  &35.73  &80.42  &7.48  &1.22    &64.73  \\
PBE+MBD    &3.423  &3.638  &3.398  &22.63  &48.96  &6.17  &0.69    &31.64  \\
PBE+dDsC   &3.447  &3.639  &3.410  &28.04  &63.00  &5.53  &0.74    &38.43  \\
ACFDT-RPA   &3.39\footnotemark[1] &  &3.34\footnotemark[2]  &   &48\footnotemark[2]  &   &      &36\footnotemark[2]   \\
Experiment  &    &         &3.34\footnotemark[3]  &  &\mcr{$43\pm5$}\footnotemark[4],
$35\pm10$\footnotemark[5], $52\pm5$\footnotemark[6]  &\mcr{7.7\footnotemark[7]}
&\mcr{0.86\footnotemark[7]} &36.5\footnotemark[8], 38.7\footnotemark[9]  \\
\hline
AIREBO      &3.391  &3.418  &3.357  &22.85  &48.86  &0.41  &0.04  &37.78  \\
LCBOP       &3.346  &3.365  &3.346  &12.51  &25.03  &0.47  &0.01  &29.77  \\
KC          &3.374  &3.602  &3.337  &21.60  &47.44  &6.07  &0.44  &34.45  \\
\ipshortname &3.439  &3.612  &3.415  &\mc{23.05}  &\mc{47.38}  &6.02  &0.71  &\mc{32.00}  \\
\end{tabular}
\end{ruledtabular}
\footnotetext[1]{Ref.~\onlinecite{zhou2015van}.}
\footnotetext[2]{Ref.~\onlinecite{lebegue2010cohesive}.}
\footnotetext[3]{Ref.~\onlinecite{baskin1955lattice}.}
\footnotetext[4]{Ref.~\onlinecite{girifalco1956energy}.}
\footnotetext[5]{Ref.~\onlinecite{benedict1998microscopic}.}
\footnotetext[6]{Ref.~\onlinecite{zacharia2004interlayer}.}
\footnotetext[7]{Ref.~\onlinecite{popov2012barriers}. \mcr{Values inferred from experimental data on shear mode frequencies.}}
\footnotetext[8]{Ref.~\onlinecite{blakslee1970elastic}.}
\footnotetext[9]{Ref.~\onlinecite{bosak2007elasticity}.}
\end{table*}

\begin{figure}
\begin{subfigure}{0.32\columnwidth}
\includegraphics[width=\columnwidth]{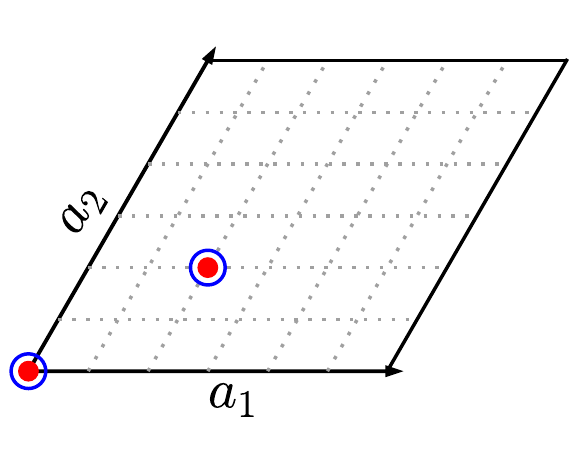}
\caption{}
\label{fig:aa}
\end{subfigure}
\begin{subfigure}{0.32\columnwidth}
\includegraphics[width=\columnwidth]{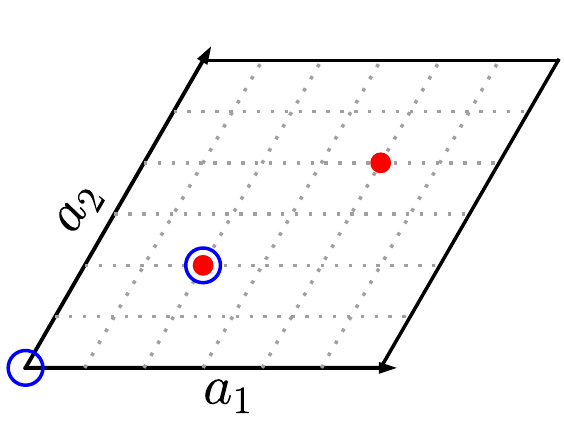}
\caption{}
\label{fig:ab}
\end{subfigure}
\begin{subfigure}{0.32\columnwidth}
\includegraphics[width=\columnwidth]{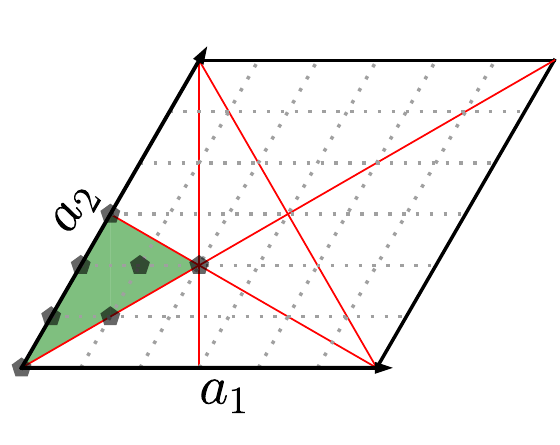}
\caption{}
\label{fig:sample:grid}
\end{subfigure}
\caption{Primitive unit cell of a graphene bilayer: (a) AA stacking, (b) AB stacking, and (c) unique sampling region and sampling points.}
\end{figure}

Each monolayer of the graphene bilayer is modeled as a slab with in-plane lattice constant $a=2.46~\text{\AA}$, and the supercell size in the direction perpendicular to the slab is set to $30~\text{\AA}$ to minimize the interaction between periodic images.
The sampling grid in reciprocal space is $20 \times 20 \times 1$, with an energy cutoff of $500~\text{eV}$.
A primitive unit cell of a graphene bilayer consists of four basis atoms.
To generate a graphene bilayer with different translational registry, the two atoms in the bottom layer are fixed at fractional positions $\bm b_1 = (0,0,0)$ and $\bm b_2 = (\frac{1}{3},\frac{1}{3},0)$ relative to the graphene lattice vectors $\bm a_1, \bm a_2$, and $\bm c$, where $\bm c$ is perpendicular to the plane defined by  $\bm a_1$ and $\bm a_2$ with length equal to the interlayer distance $d$.
The other two atoms are located at $\bm r_1 = (p,q,1)$ and  $\bm r_2 = (p+\frac{1}{3}, q+\frac{1}{3}, 1)$.
The two parameters $p\in[0,1]$  and $q\in[0,1]$ determine the translational registry.
For example, the graphene bilayer is in AA stacking (\fref{fig:aa}) when $p=0$ and $q=0$, and in AB stacking (\fref{fig:ab}) when $p=\frac{1}{3}$ and $q=\frac{1}{3}$.
Due to the symmetry of the honeycomb lattice, only 1/12 of the area defined by $\bm a_1$ and $\bm a_2$ needs to be sampled to fully explore all translational
registry states (see the shaded region in \fref{fig:sample:grid}).
The \ipshortname training set comprised the seven states indicated in the
shaded region of \fref{fig:sample:grid}, specifically
$(p,q)=(0,0)$, $(0,\frac{1}{6})$, $(0,\frac{2}{6})$, $(0,\frac{3}{6})$,
$(\frac{1}{6}, \frac{1}{6})$, $(\frac{1}{6}, \frac{2}{6})$,
$(\frac{2}{6}, \frac{2}{6})$.
\mc{These states include all the high-symmetry states of interest, including AA, AB, and the saddle point (SP) stacking ($p=0, q=\frac{3}{6}$).
The seven translational registry states are sampled at different layer spacings
$d$, varying from $2.7~\text{\AA}$ to $4.5~\text{\AA}$ with a step size of
$0.1~\text{\AA}$.
For layer spacings larger than $4.5~\text{\AA}$ but smaller than the cutoff
$r_\text{cut} = 12~\text{\AA}$, only bilayer graphene in AB stacking is included
since the difference between the stacking states in this range is negligible
(see discussion in \sref{sec:test}).
Thus $7\times 19 + 75 = 208$ translation configurations are included in the training set.
}

\begin{figure}
\begin{subfigure}{0.6\columnwidth}
\includegraphics[width=\columnwidth]{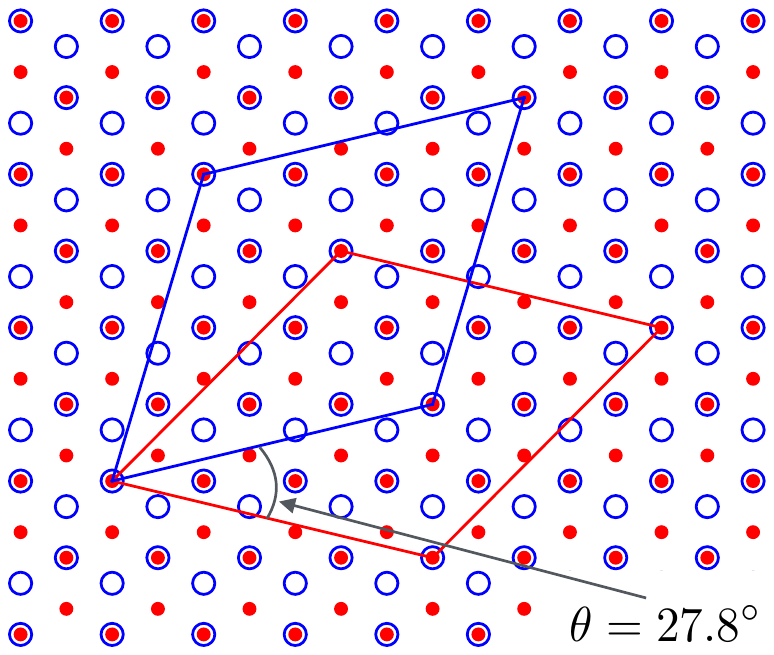}
\caption{}
\label{fig:twisted:bilayer:before}
\end{subfigure}
\begin{subfigure}{0.35\columnwidth}
\includegraphics[width=\columnwidth]{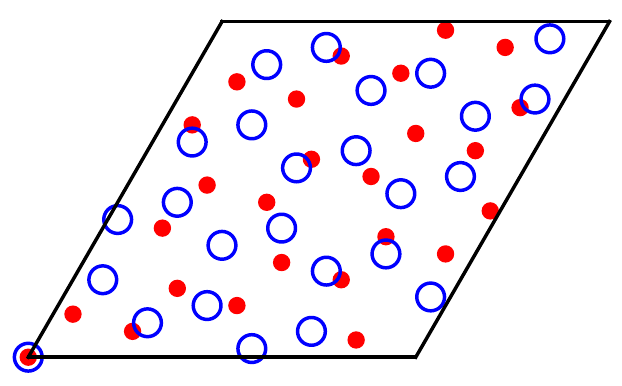}
\caption{}
\label{fig:twisted:bilayer:after}
\end{subfigure}
\caption{Example of commensuration of a graphene bilayer. (a) The two layers are commensurate when rotated relative to each other by $\cos^{-1} (\frac{23}{26})=27.8^\circ$, which corresponds to $m=3, n=7$ according to the condition in
\eqn{eq:commensuration}. (b) The resulting supercell after rotation, with 26 atoms in each layer.}
\label{fig:twisted:bilayer}
\end{figure}

In addition to translation configurations, a set of
twisted bilayer configurations are included in the training set.
It is possible to construct a commensurate supercell arbitrarily close to any twisting angle according to the commensuration condition\cite{shallcross2010electronic,zhang2017energy,tritsaris2016}
\begin{equation}\label{eq:commensuration}
\theta = \cos^{-1}\left(\frac{3n^2 - m^2}{3n^2 + m^2}\right) ,
\end{equation}
where $m$ and $n$ are any two integers satisfying $0 < m < n$.
As an example, considering the AB-stacked bilayer in \fref{fig:twisted:bilayer:before}, a commensurate bilayer can be obtained by rotating one of the layers by  $\theta = 27.8^\circ$ ($m=3,n=7$) with the supercell shown in \fref{fig:twisted:bilayer:after}.
Four types of twisted bilayers with rotation angles $9.43^\circ$, $21.79^\circ$, $32.30^\circ$ and $42.10^\circ$ (corresponding to $(m,n) = (1,7)$, $(1,3)$, $(1,2)$ and $(2,3)$) are included in the training set.
\mc{The twisted configurations were evaluated at layer spacings from 3.0~\AA\xspace
 to 4.0~\AA\xspace with a step size of 0.1~\AA.
Thus $4 \times 11 = 44$ twisted configurations are included in the training set.
}
This does not include rotations for
$\theta = 0^\circ$ and $\theta = \pm60^\circ$ corresponding
to the AB and AA stacking states, respectively,
which are already included in the training set.

The parameters of the potential are optimized by minimizing a loss function that quantifies the difference between the interatomic potential predictions and the training set. The training set includes $M$ configurations with
concatenated coordinates $\bm r_m$ for $m\in[1,M]$, such that
$\bm r_m\in\mathbb{R}^{3N_m}$ where $N_m$ is the number of atoms
in configuration $m$.  The loss function is
\begin{align} \label{eq:loss}
L(\bm{\xi})
=& \sum_{m=1}^M \frac{1}{2} w^\text{e}_m \left[E_m(\bm r_m; \bm{\xi}) - E_m^\text{DFT} \right]^2 \nonumber \\
+& \sum_{m=1}^M \frac{1}{2} w^\text{f}_m \|\bm f(\bm r_m; \bm{\xi}) - \bm f_m^\text{DFT} \|^2,
\end{align}
where $\bm{\xi}$ is the set of potential parameters,
$E_m$ and
$\bm{f}(\bm r_m; \bm{\xi}) = -\left.(\partial\mathcal{V}/\partial \bm{r})\right|_{\bm{r}_m} \in \mathbb{R}^{3N_m}$
are the \ipshortname potential energy and concatenated forces
in configuration $m$,
and $w^\text{e}_m$ and $w^\text{f}_m$ are the weights associated with
the energy and forces of configuration $m$. For energy in units of eV
and forces in units of eV/\AA, these weights
have units of eV$^{-2}$ and (eV/\AA)$^{-2}$, respectively.

The DFT energy and forces used in the loss function, \eqn{eq:loss},
$E_m^\text{DFT}$ and $\bm f_m^\text{DFT}$, require explanation.
Since DFT provides only the total energy and forces on atoms
due to both intralayer and interlayer interactions it is necessary
to separate out the interlayer contributions when constructing the training set.
This is accomplished as follows. For configuration $m$,
first the total energy and forces of the bilayer are
obtained from DFT: $E_m^\text{DFT, bilayer}$, $\bm f_m^\text{DFT, bilayer}$.
Then each monolayer is computed separately by
removing all atoms from the other monolayer. Thus, there will be two
energies, $E_m^\text{DFT, layer 1}$ and $E_m^\text{DFT, layer 2}$,
and two forces, $\bm f_m^\text{DFT, layer 1}$ and $\bm f_m^\text{DFT, layer 2}$
(although each force vector will only contain nonzero components for the
atoms belonging to its monolayer).
The DFT interlayer energy and forces appearing in \eqn{eq:loss}
are then defined as:
\begin{align}
E_m^\text{DFT} &= E_m^\text{DFT, bilayer}
- E_m^\text{DFT, layer 1} - E_m^\text{DFT, layer 2},
\label{eq:E:DFT}
\\
\bm f_m^\text{DFT} &=
\bm f_m^\text{DFT, bilayer}
- \bm f_m^\text{DFT, layer 1} - \bm f_m^\text{DFT, layer 2}.
\label{eq:f:DFT}
\end{align}

In the present case, the training set includes $M=252$ configurations.
Both the energy weight $w^\text{e}_m$ and force weight $w^\text{f}_m$
($m=1,\dots,252$) are set to 1.
The optimization was carried out using the KIM-based Learning-Integrated
Fitting Framework (KLIFF)\cite{kliff}
with a geodesic Levenberg-Marquardt minimization algorithm\cite{wen2016potfit, transtrum2010nonlinear,transtrum2011geometry}.
The objective is to find the set of parameters $\bm{\xi}$ that minimizes $L(\bm{\xi})$.
The optimal parameter set identified by this process and preset cutoffs
are listed in \tref{tab:params}.

\begin{table}
\caption{\mc{\ipshortname parameters obtained by minimizing the loss function
$L(\bm\xi)$ defined in \eref{eq:loss} and preset cutoffs.}}
\label{tab:params}
\begin{ruledtabular}
\begin{tabular}{cccc}
 Parameter        &Value    &Parameter              &Value  \\
\hline
 $C_0$~(meV)       &11.598   &$B$~(meV)               &7.6799 \\
 $C_2$~(meV)       &12.981   &$\eta$~(1/\AA)          &1.1432 \\
 $C_4$~(meV)       &32.515   &A~(meV)                 &22.216 \\
 $C$~(meV)         &7.8151   &$z_0$~(\AA)             &3.3400 \\
 $\delta$~(\AA)    &0.83679  &$r_\text{cut}$~(\AA)    &\mc{12} \\
 $\lambda$~(1/\AA) &2.7158   &$\rho_\text{cut}$~(\AA) &1.562  \\
\end{tabular}
\end{ruledtabular}
\end{table}

\section{Testing of the new potential}
\label{sec:test}
We performed an extensive set of calculations to test
the ability of \ipshortname to reproduce its training set
(described in \sref{sec:model}),
and test its transferability to configurations outside the training set.
The calculations using the potential were performed with LAMMPS\cite{plimpton1995fast,lammps} and DFT calculations with VASP\cite{vasp1,vasp2}.
Periodic boundary conditions are applied in both in-plane directions, and the in-plane lattice constant is fixed at $a=2.46~\text{\AA}$.
The setup for the DFT computations is the same as that used for generating the training set in \sref{sec:model}.

\begin{figure*}
\begin{subfigure}{0.65\columnwidth}
\includegraphics[width=\columnwidth]{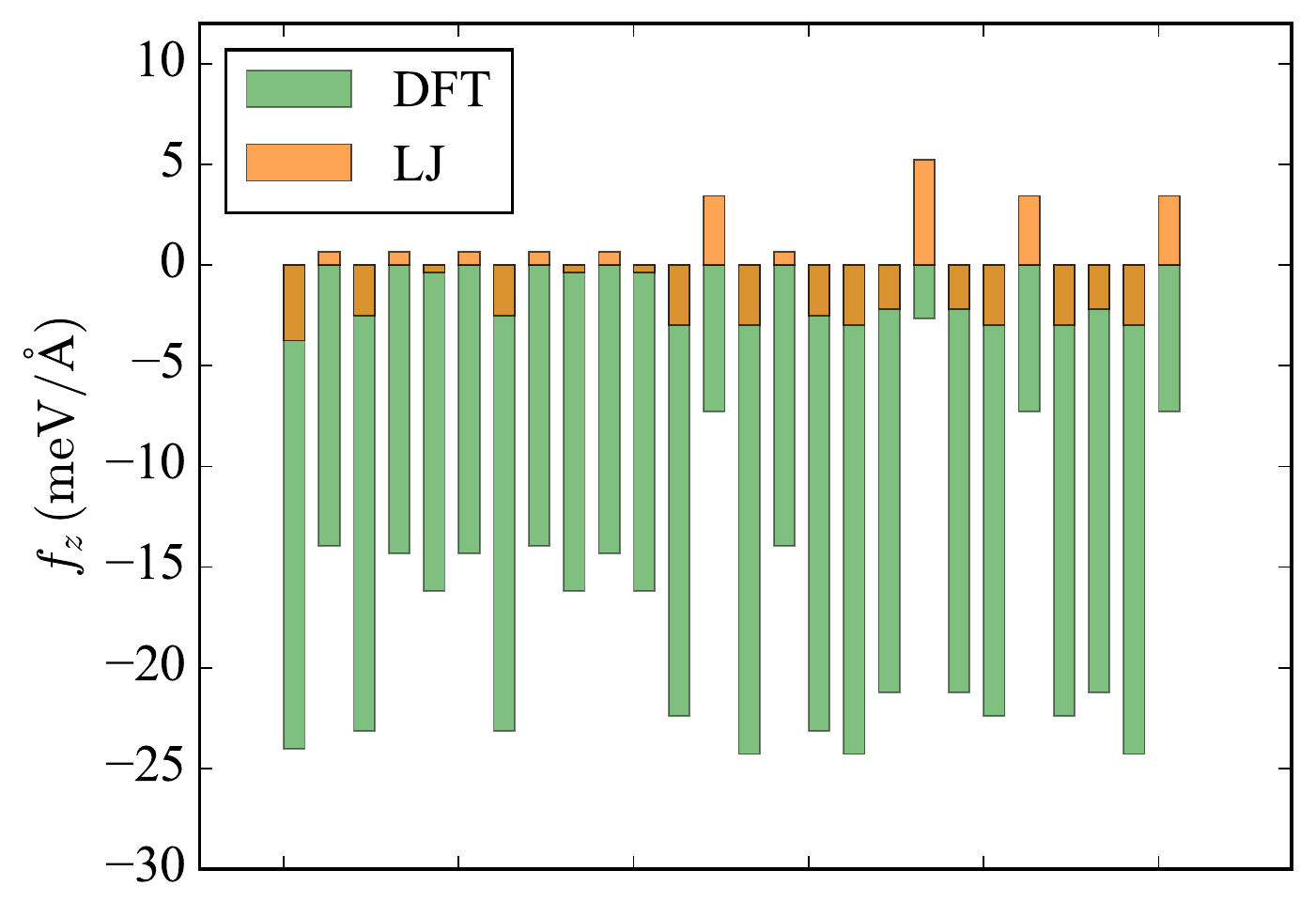}
\caption{}
\label{fig:z:forces:lj}
\end{subfigure}
\begin{subfigure}{0.65\columnwidth}
\includegraphics[width=\columnwidth]{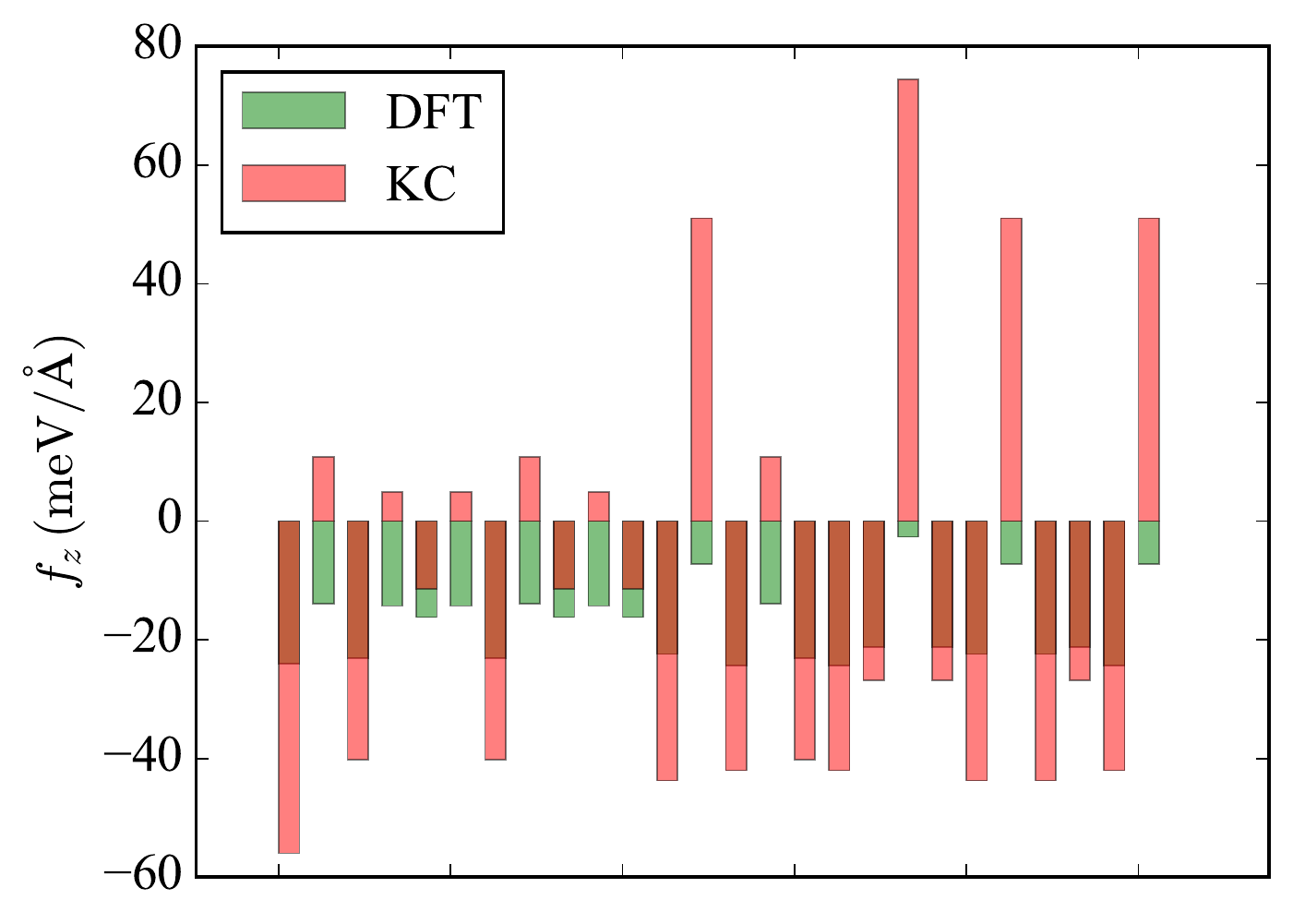}
\caption{}
\end{subfigure}
\begin{subfigure}{0.65\columnwidth}
\includegraphics[width=\columnwidth]{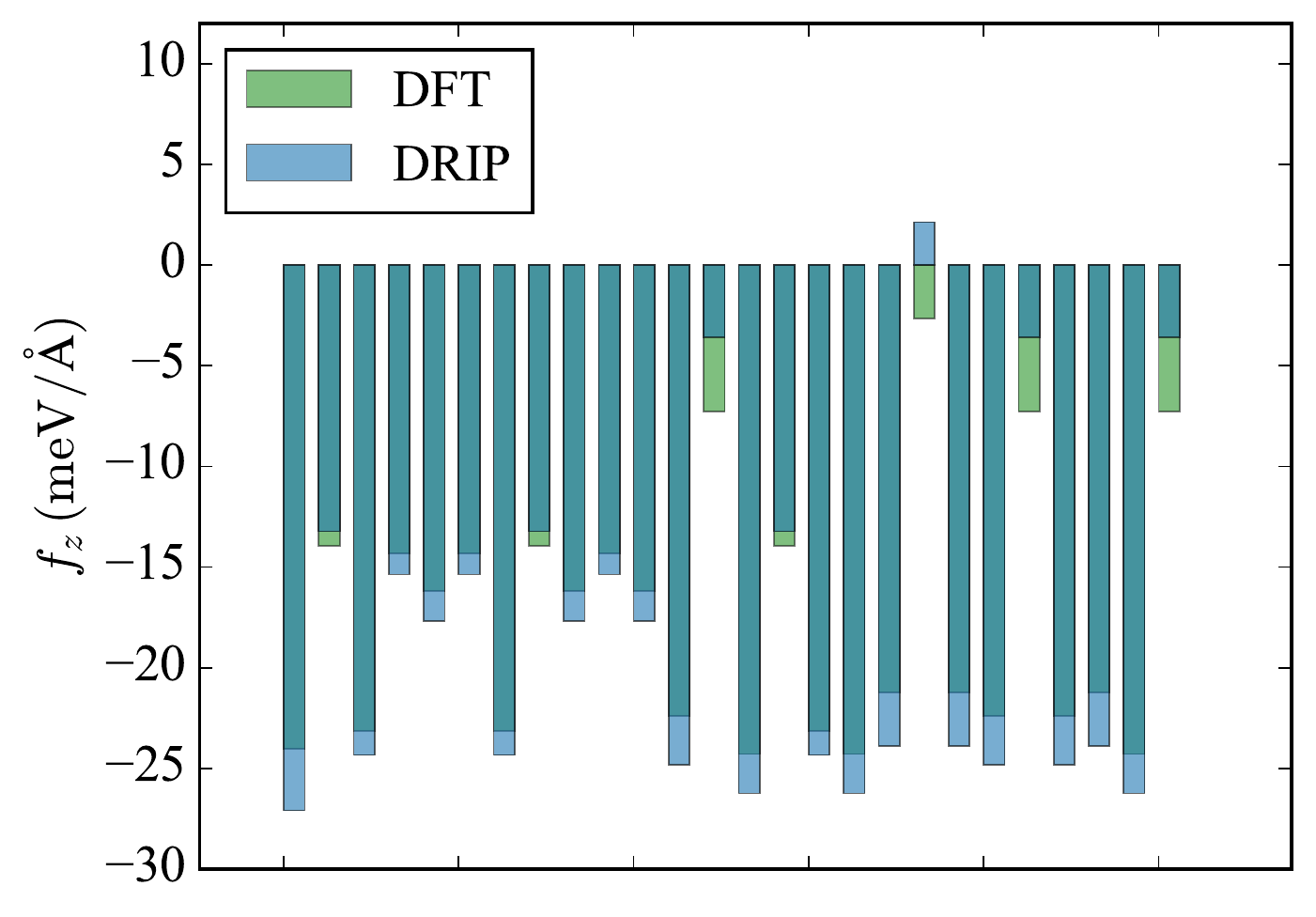}
\caption{}
\end{subfigure}
\caption{Out-of-plane component of the forces on the 26 atoms in the bottom layer of the twisted bilayer shown in \fref{fig:twisted:bilayer} (each represented as a bar) computed from DFT and the (a) LJ potential, (b) KC potential, and (c) \ipshortname model.  The layer spacing is 3.4~\AA.}
\label{fig:z:forces}
\end{figure*}

\fref{fig:z:forces} shows the unrelaxed forces on the atoms in the bottom layer of the twisted bilayer shown in \fref{fig:twisted:bilayer} with a layer spacing of $d=3.4~\text{\AA}$.
There are 26 atoms in the bottom layer. For each, the out-of-plane force ($z$-component)
 is displayed as a bar.
The plot compares the results of LJ, KC and \ipshortname with DFT.
For the LJ potential, the parameterization in the AIREBO potential is used.
The \ipshortname forces are in very good agreement with DFT, whereas the LJ
potential yields almost zero forces, and the KC potential greatly overestimates
the forces. (Note that the force ranges in the three panels are different).
The force on the central atom when twisting a bilayer obtained from \ipshortname (denoted as 1 in \fref{fig:high:sym:confs}) is
displayed in \fref{fig:rotation:force} as a function of rotation.
The results are in agreement with DFT, indicating that the
dihedral modification in \ipshortname successfully addresses the deficiency
of the KC potential discussed in \sref{sec:need}.

\begin{figure}
\includegraphics[width=1\columnwidth]{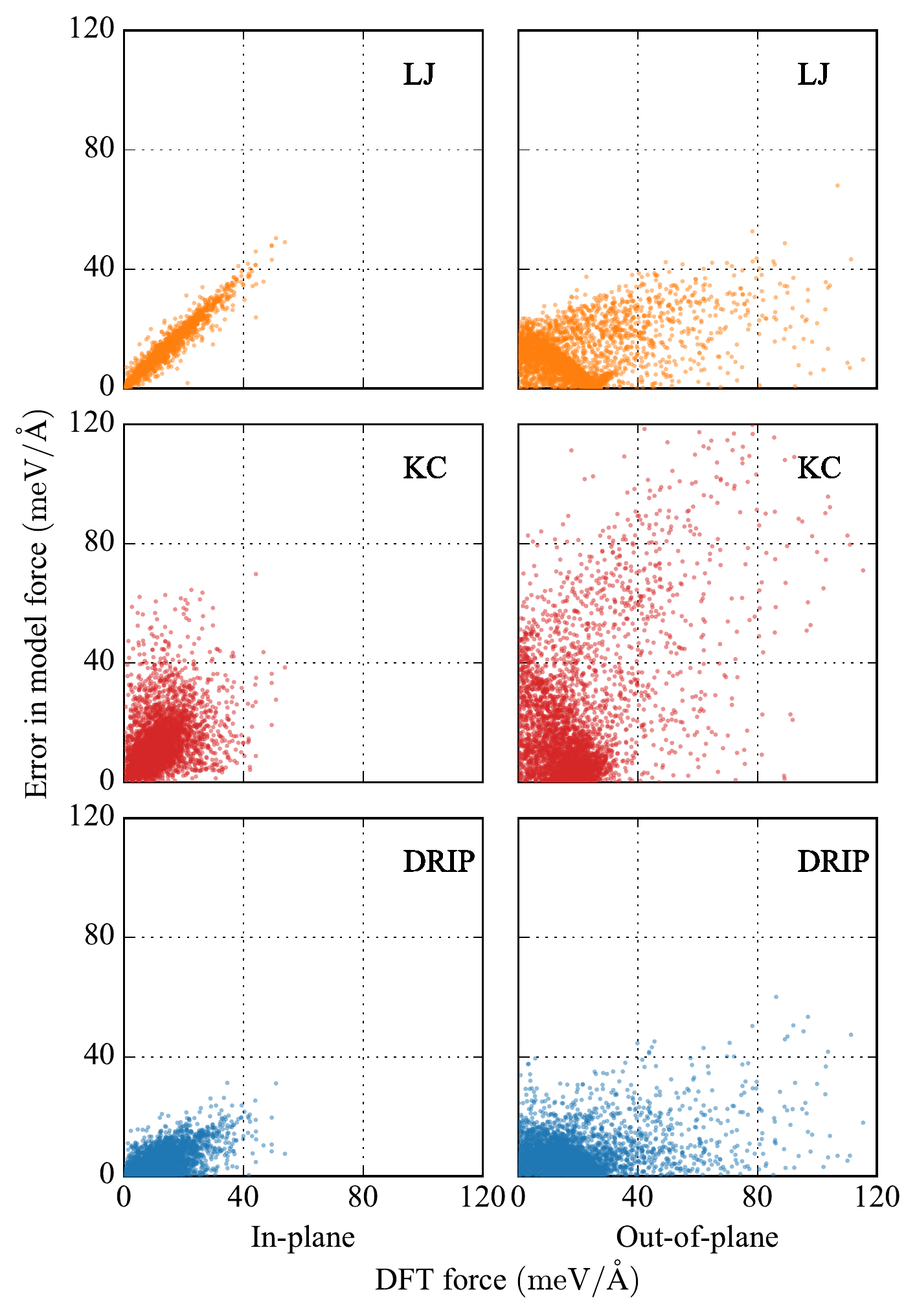}
\caption{\mc{Deviation of potential forces from DFT results due to interlayer
interactions. The configurations are taken from three AIMD trajectories at 300~K.}}
\label{fig:force:error}
\end{figure}

To investigate the accuracy of the potentials in a dynamical setting,
trajectories are generated at a temperature of 300~K
using \emph{ab initio} molecular dynamics (AIMD)
for bilayers in AA and AB stackings, and the twisted bilayer shown in
\fref{fig:twisted:bilayer}.
For each configuration along the trajectories, the DFT forces due to interlayer
interactions are computed using the procedure defined in \eqn{eq:f:DFT}
and explained above.  Next, LAMMPS is used to compute the LJ, KC and \ipshortname
interlayer forces for the AIMD configurations.
The error in the potential forces is shown in \fref{fig:force:error}.
Each dot in the plot represents one atom pulled from one of the configurations
along the AIMD trajectories.
\mc{The horizontal coordinate in the plot is the magnitude of the in-plane component (left panels)
and out-of-plane component (right panels) of the DFT interlayer force acting on the atom. The force
is separated in this way because the in-plane component is significantly smaller than the out-of-plane
component. (Note that this is only the force due to interlayer interactions. The force due
to intralayer bonding is not included.)
The vertical coordinate is the magnitude of the
difference between the potential and DFT force vectors for that atom.
We see that the in-plane force error for LJ aligns with the diagonal,
i.e.\ the error equals the DFT force, which means that LJ predicts an
in-plane force close to zero.
This is because LJ provides a poor model for the
anisotropic overlap of electronic orbitals between adjacent layers and thus
has almost no barrier for relative sliding.
The KC model performs better in the sense that it predicts resistance
to sliding, however the overall accuracy in forces is poor
(see \sref{sec:need} for a discussion of the limitations of the KC model).
In contrast, \ipshortname provides consistently accurate in-plane forces
across the range of DFT forces with errors less than 20~meV/\AA.
For the out-of-plane component both LJ and \ipshortname perform comparably
providing good accuracy across the range of DFT forces, whereas
the KC model again shows poor accuracy with very large errors
in some cases.}

\begin{figure*}
\begin{subfigure}{0.65\columnwidth}
\includegraphics[width=\columnwidth]{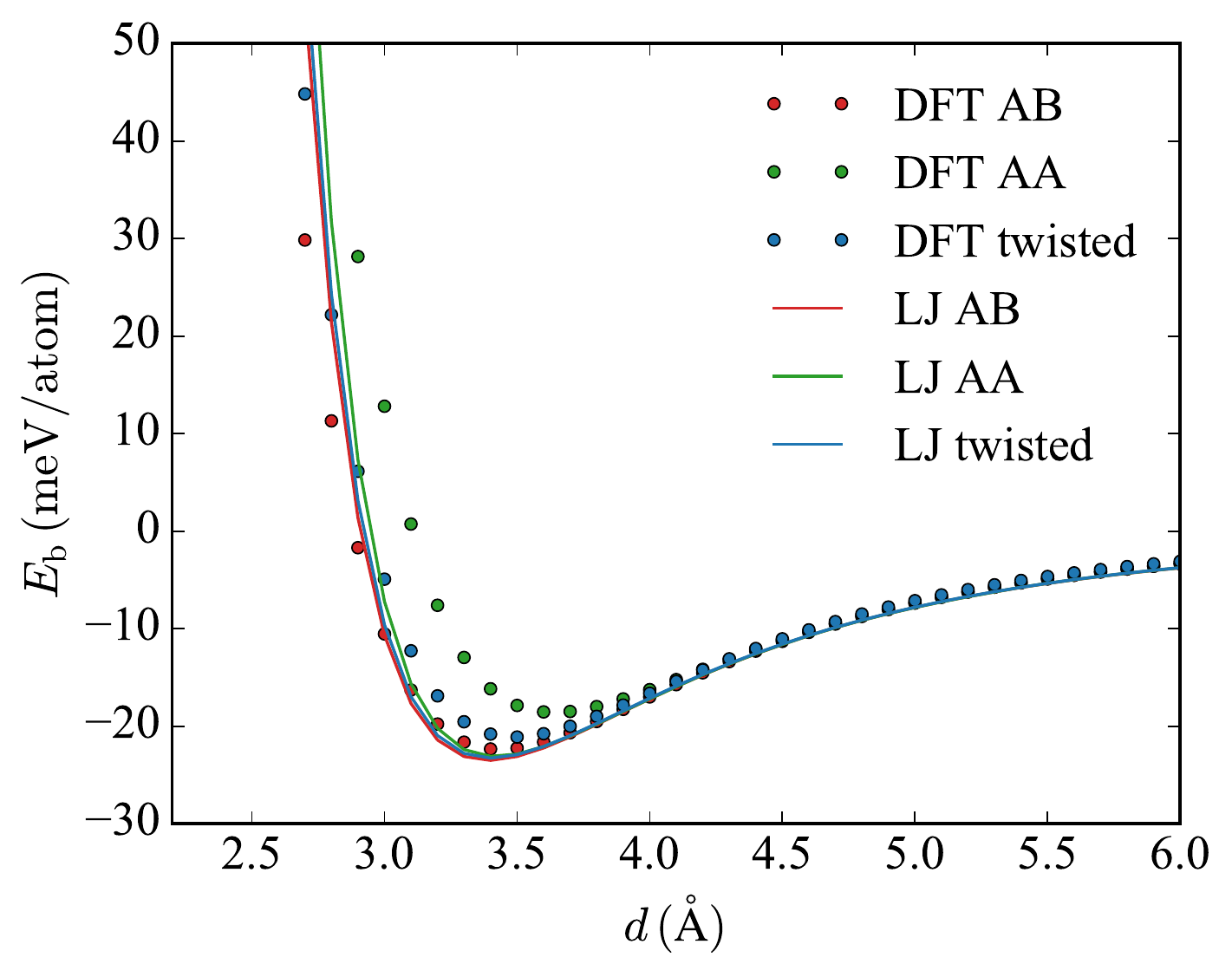}
\caption{}
\label{fig:energy:layer:dist:lj}
\end{subfigure}
\begin{subfigure}{0.65\columnwidth}
\includegraphics[width=\columnwidth]{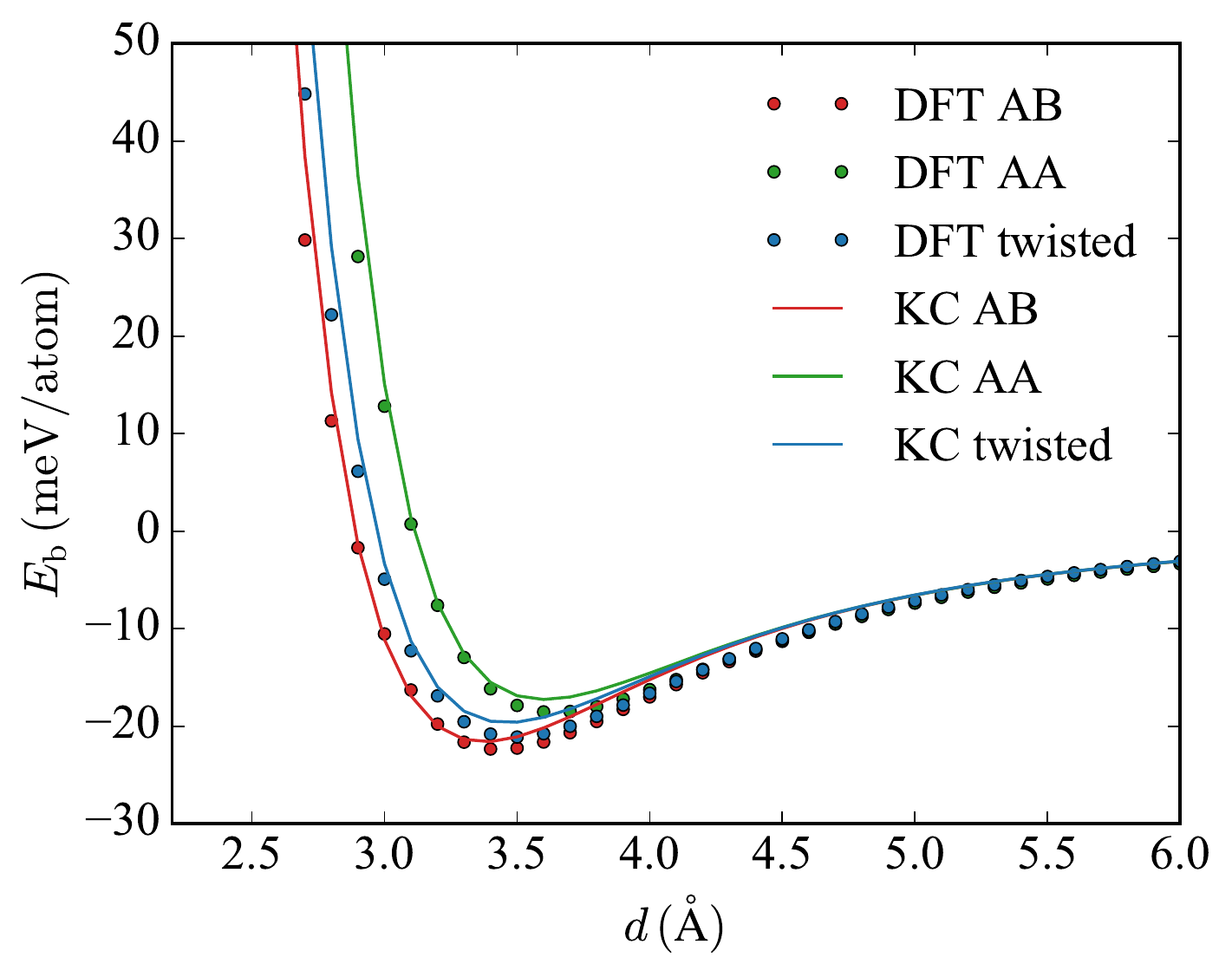}
\caption{}
\label{fig:energy:layer:dist:kc}
\end{subfigure}
\begin{subfigure}{0.65\columnwidth}
\includegraphics[width=\columnwidth]{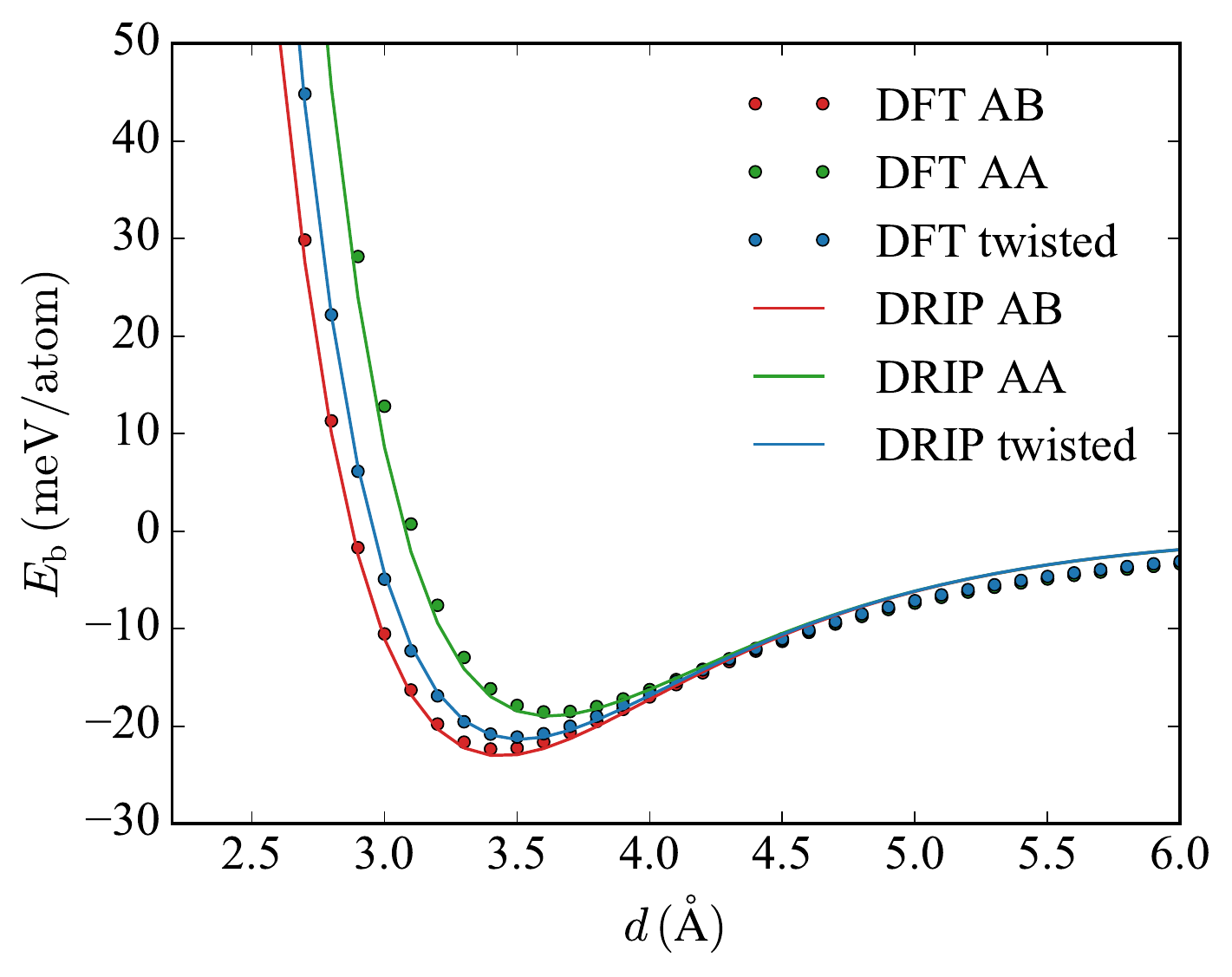}
\caption{}
\label{fig:energy:layer:dist:present}
\end{subfigure}
\caption{Interlayer binding energy $E_\text{b}$ of a graphene bilayer versus layer spacing $d$ for AA stacking, AB stacking, and a twisted bilayer with rotation angle $\theta=27.8^\circ$ (see \fref{fig:twisted:bilayer}) using (a) LJ potential, (b) KC potential, and (c) \ipshortname model, compared to DFT results.}
\label{fig:energy:layer:dist}
\end{figure*}

Next, we consider energetics.
The interlayer binding energy $E_\text{b}$ of a graphene bilayer as a function of layer spacing $d$ is shown in \fref{fig:energy:layer:dist} for AB and AA stackings and the twisted configuration shown in \fref{fig:twisted:bilayer}.
The LJ potential (\fref{fig:energy:layer:dist:lj})
cannot distinguish these states and gives nearly identical binding energy versus layer spacing curves for all three.
Both KC (\fref{fig:energy:layer:dist:kc}) and \ipshortname (\fref{fig:energy:layer:dist:present}) correctly capture the energy differences between the three stacking states.
For all three potentials, the twisted bilayer curve lies between the other two,
which is expected since the AB and AA stackings are minimum and maximum energy
states. Also notable is that at large layer spacing, the curves for all three
stacking states merge since registry effects due to $\pi$-orbital overlap
become negligible and interactions are dominated by vdW attraction, which are the
same for all three states and captured equally well by all three potentials.

\begin{figure}
\includegraphics[width=\columnwidth]{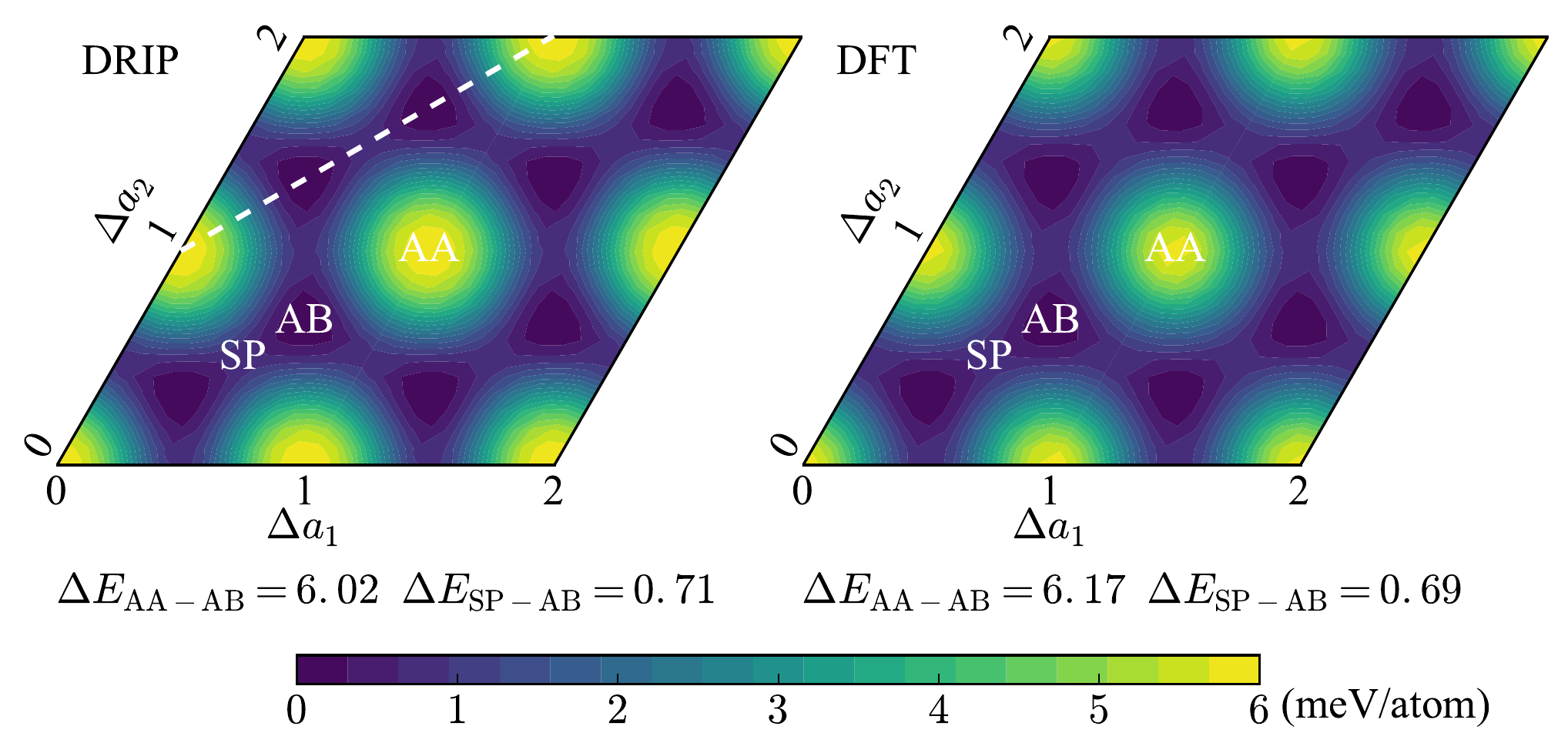}
\caption{
  \mc{The GSFE obtained by sliding one layer relative to the other at a fixed
  layer spacing of $d=3.4~\text{\AA}$.
  The energy is relative to the AB state, which is $-22.98~\text{meV/atom}$
  for \ipshortname (on the left) and $-22.33~\text{meV/atom}$ for DFT (on the
  right).
  The sliding parameters $\Delta \bm a_1$ and $\Delta \bm a_2$ are in units of
  in-plane lattice constant $a=2.46~\text{\AA}$.
  }
}
\label{fig:gsfe}
\end{figure}

A more complete view of the interlayer energetics is obtained by considering
the generalized stacking fault energy (GSFE) surface obtained by sliding one layer
relative to the other while keeping the layer spacing fixed.
\fref{fig:gsfe} shows the results for a layer spacing of $d=3.4~\text{\AA}$
calculated using \ipshortname and DFT.
\ipshortname is in quantitative agreement with DFT results.
The KC GSFE has a similar appearance and the LJ GSFE is nearly flat.
The KC and LJ results are not included for
brevity, but the energies of the three potentials along the dashed line in the
left panel of \fref{fig:gsfe} are displayed in \fref{fig:sliding:energy}.

As a final test, \tref{tab:properties} shows the predictions of \ipshortname for
a number of structural, energetic, and elastic properties. The table also
includes results for the LCBOP\cite{los2003intrinsic} and
AIREBO\cite{stuart2000airebo} potentials, as well as DFT and experimental
results as described in \sref{sec:model}.
The LCBOP potential uses two Morse\cite{morse1929diatomic} type terms to
model long-range interactions, and LJ\cite{lennardjones1931}
is used in the AIREBO potential as discussed in \sref{sec:intro}.
The properties of the \ipshortname model are in good agreement with the PBE+MBD
DFT computations with which the training set was generated.


\section{Applications}
\label{sec:app}

To further compare the predictions of the KC potential and \ipshortname,
we carried out
two large-scale simulations, beyond the capability of DFT:
(1) structural relaxation in a twisted graphene bilayer, and (2) exfoliation of a graphene layer off graphite. In these simulations, the interlayer interactions are modeled using either KC or \ipshortname, and the REBO\cite{brenner2002rebo} potential is used to model the intralayer interactions.

\subsection{Structural relaxation of a twisted graphene bilayer}
\label{sec:bilayer:relax}

The electronic properties of stacked 2D materials can be manipulated by controlling
the relative rotation between the layers, which in turn leads to different
structural relaxation. A prototypical problem is the twisting of a graphene bilayer.
The bilayer is created by rotating one layer relative to the other by $\theta=0.82^\circ$,
setting $(m,n) = (1,81)$ as discussed in \sref{sec:model}.
The out-of-plane relaxation
$\delta$ of an atom is obtained by subtracting the mean out-of-plane coordinates of all atoms in the top layer from the out-of-plane coordinate of that atom:
\begin{equation}
\delta_i = z_i - \frac{1}{N} \sum_{j=1}^N z_j
\end{equation}
where $z_i$ is the out-of-plane coordinate of atom $i$ in the top layer and $N=9842$ is the number of atoms in the top layer\footnote{Using the atoms in the bottom layer will yield the same results because the relaxed structure of the bottom layer and the top layer are identical.}.

\begin{figure}
\begin{subfigure}{1.0\columnwidth}
\includegraphics[width=\columnwidth]{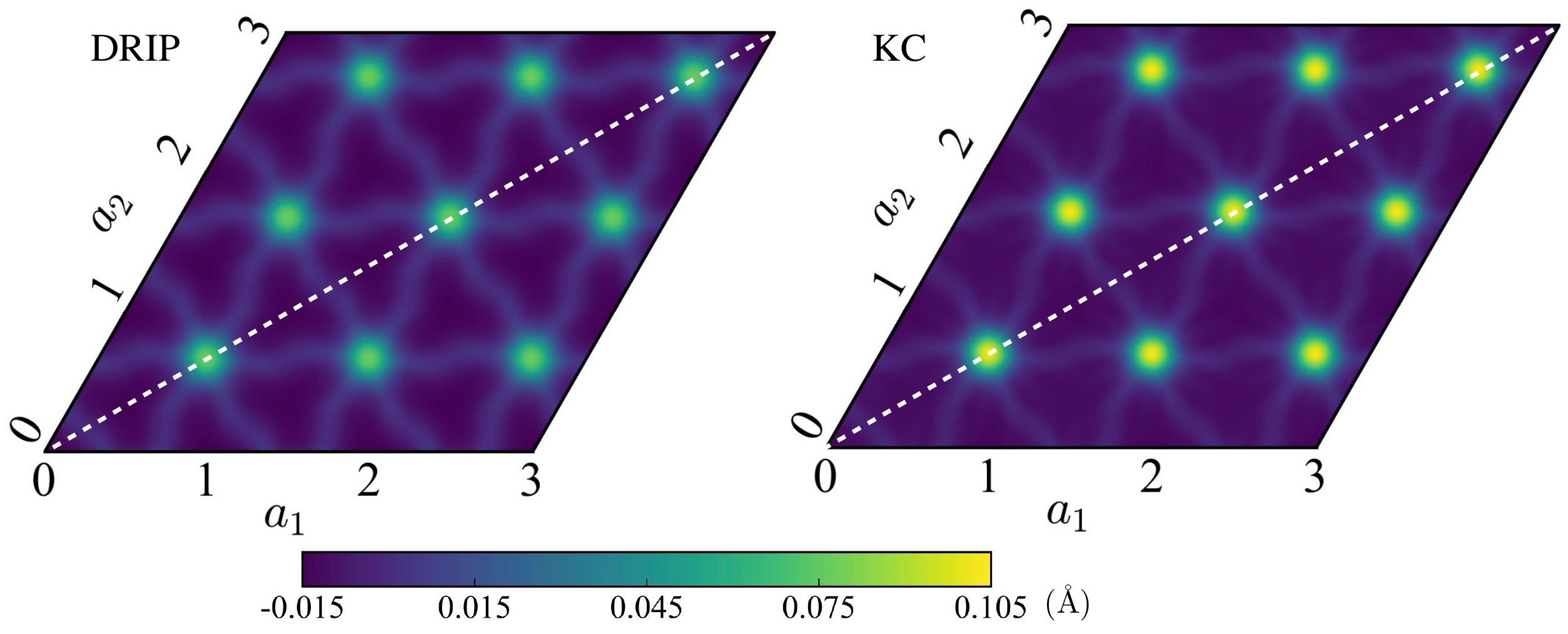}
\caption{}
\label{fig:twisted:bilayer:relax:a}
\end{subfigure}
\begin{subfigure}{0.9\columnwidth}
\includegraphics[width=\columnwidth]{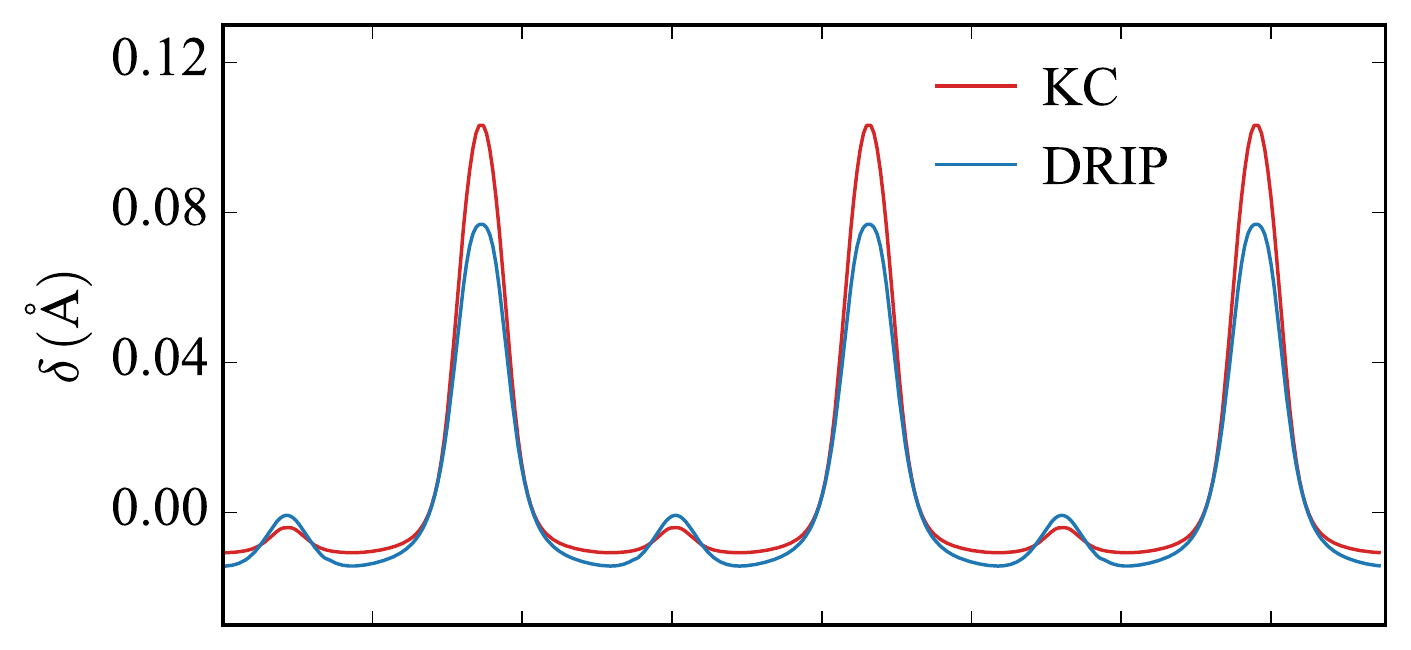}
\caption{}
\label{fig:twisted:bilayer:relax:b}
\end{subfigure}
\caption{Out-of-plane relaxation in a twisted bilayer with a relative rotation of $\theta=0.82^\circ$.
(a) Contour plot obtained from the \ipshortname model and the KC potential, and (b) relaxation along the diagonal indicated by the dashed line in panel (a).
\mc{The bilayers shown in the figure corresponds to $3\times3$ supercells used in the computation,
i.e.\  $a_1$ and $a_2$ are in units of in-plane lattice constant $a=2.46~\text{\AA}$.} }
\label{fig:twisted:bilayer:relax}
\end{figure}

The out-of-plane relaxation of the twisted bilayer is plotted in \fref{fig:twisted:bilayer:relax}.
The results of the \ipshortname and KC models are qualitatively similar.
The bright spots correspond to high-energy AA stacking, the long
narrow ribbons correspond to SP stacking, and the triangular regions
correspond to alternating AB and BA stacking. It has been shown that the
formation of this structure is due to local rotation at AA domains.\cite{zhang:tadmor:2018}
Quantitatively, however, the two potentials give different out-of-plane relaxation, especially at the peaks as seen in \fref{fig:twisted:bilayer:relax:b}.
The peak value predicted by \ipshortname is $0.076~\text{\AA}$, which is $26\%$ smaller than the KC potential value of $0.103~\text{\AA}$.
This difference at the peaks could lead to significant differences in electronic properties because twisted graphene bilayers develop highly-localized states around AA-stacked regions for small twist angles
\cite{gonzalez2017electrically}.

\subsection{Exfoliation of graphene from graphite}

\begin{figure}
\begin{subfigure}{0.9\columnwidth}
\includegraphics[width=\columnwidth]{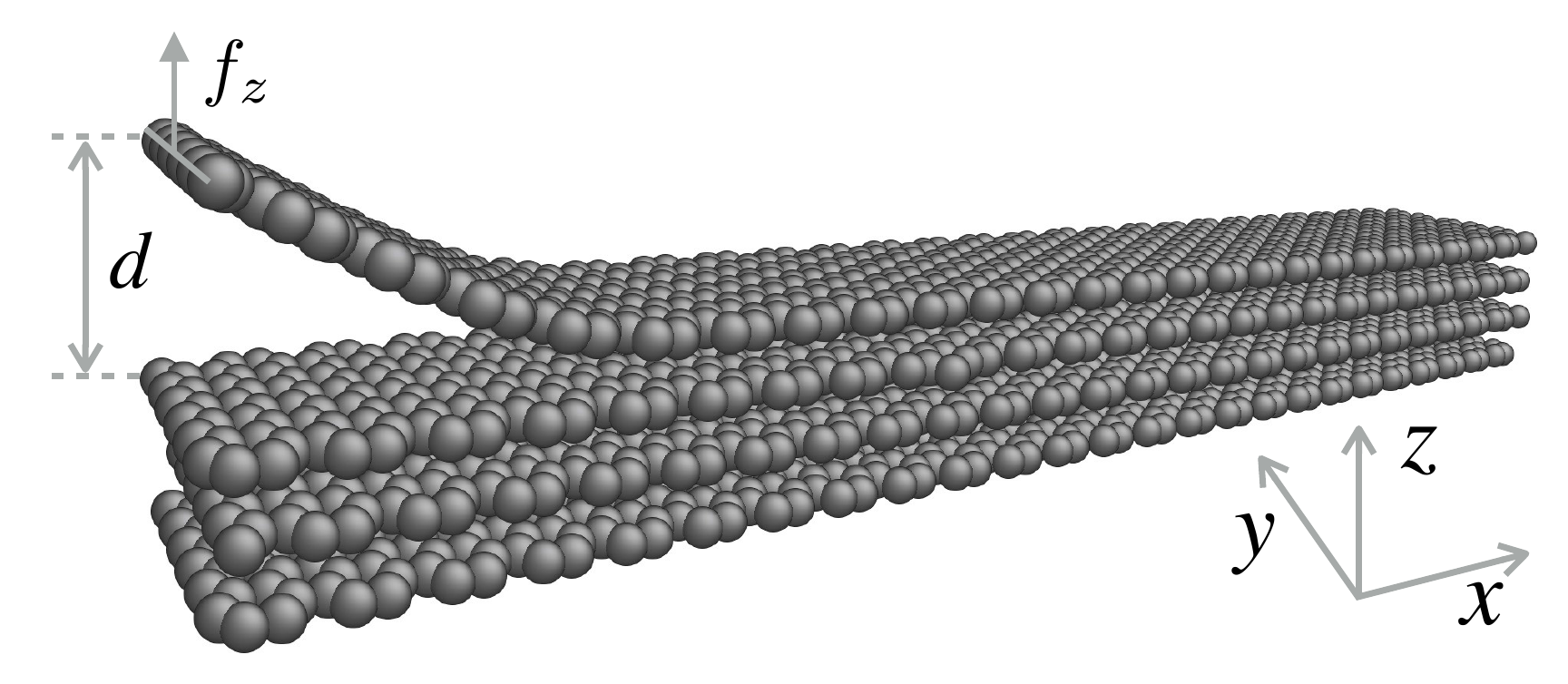}
\caption{}
\label{fig:peel:one:layer:a}
\end{subfigure}
\begin{subfigure}{0.9\columnwidth}
\includegraphics[width=\columnwidth]{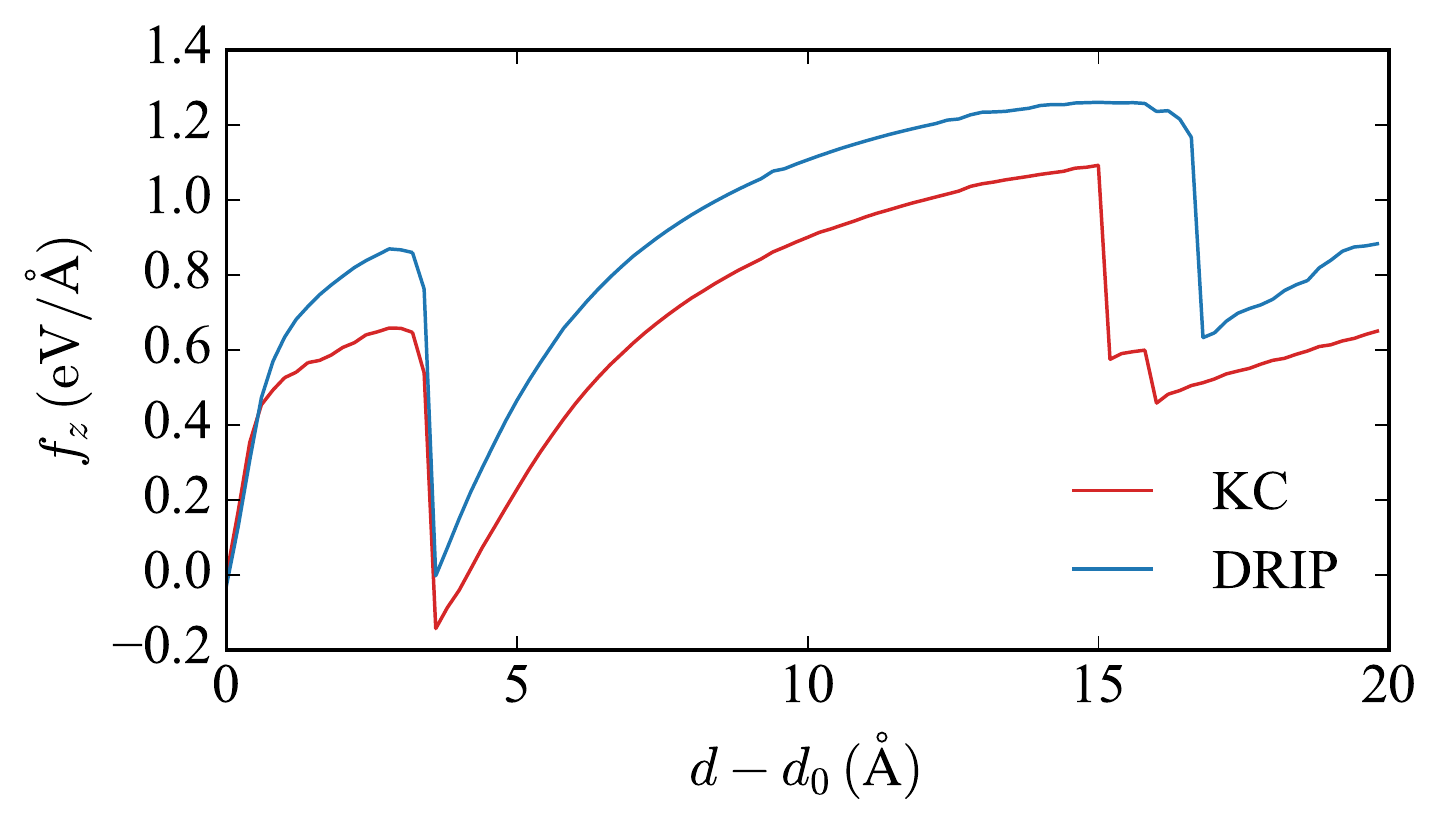}
\caption{}
\label{fig:peel:one:layer:b}
\end{subfigure}
\caption{(a) Schematic demonstrating the process of peeling a graphene layer off graphite, and (b) the normal force, $f_z$, needed to peel the top layer as a function of the displacement at the left end of the top layer, $d - d_0$. The armchair direction of
graphite is aligned with the $x$-axis. The initial layer spacing is $d_0=3.35~\text{\AA}$.}
\label{fig:peel:one:layer}
\end{figure}

Graphene can be prepared by exfoliating graphite. In this process, the vdW
attraction between layers is overcome by peeling a single layer off a
graphite crystal. A method as simple as sticking scotch tape to graphite and
applying an upward force can be used.\cite{novoselov2004electric}
To simulate this process, one edge of the top layer of a graphite crystal
is pulled up under displacement control conditions as illustrated in
\fref{fig:peel:one:layer:a}.
The atoms at the left end of the top layer are displaced in the $z$-direction according to $d = d_0 + 0.2k$, where $d_0=3.35~\text{\AA}$ is the initial layer spacing, and $k = 0, 1, \dots, 99$ is the step number.
At each step $k$, once the displacement is applied to the left atoms,
the remaining atoms in the top layer are relaxed.
The substrate (bottom three layers) is kept rigid during this process.
The system contains 600 atoms in each layer of size $105.83~\text{\AA}$ and $14.76~\text{\AA}$ in the $x$ and $y$ directions, respectively.
The system is periodic in the $y$ direction, and non-periodic the other two directions.

The normal force, $f_z$, needed to pull the left end of the top layer is plotted in \fref{fig:peel:one:layer:b}.
Both the KC and \ipshortname models give qualitatively similar results.
The force first increases as the left end is pulled up and then exhibits a sudden
drop at about 3~\AA. The normal force has two contributions:
(1) interlayer interactions with atoms in the substrate;
and (2) covalent-bonded interactions with other atoms in the top layer.
The former is almost unchanged before and after the load drop, therefore the drop is mainly due to the in-plane interactions in the top layer.
Before the load drop, the right-end of the top layer is trapped in a local minimum created by the substrate (similar to the one denoted as AB in \fref{fig:gsfe}, although there we only consider a graphene bilayer), and consequently as the left end is pulled up, the top layer experiences an increasing axial strain.
At about $3~\text{\AA}$, the right-end of the top layer snaps into an adjacent local minimum by moving in the negative $x$ direction (see Supplemental Material for a movie showing the snap-throughs associated with the load drop).
As a result, the axial strain in the top layer is released and the load is
reduced.
The same explanation applies to the load drop at a displacement of about $15~\text{\AA}$, and it is expected to continue to occur periodically
with continued pulling.

As for the results in \sref{sec:bilayer:relax}, KC and \ipshortname are in
qualitative agreement, but there are quantitative differences.
The KC potential predicts an initial peeling load of about 0.65~eV/\AA,
which is about 75\% of the 0.87~eV/\AA\xspace value predicted by \ipshortname.
The second snap-through occurs at a displacement of $16.6~\text{\AA}$ for \ipshortname,
and at $15.0~\text{\AA}$ for KC.

\section{Summary}
\label{sec:sum}

The interlayer interactions in stacked 2D materials play an important role in determining the functionality of many nanodevices.
For graphitic structures, the two-body pairwise LJ potential is too smooth to model the energy corrugation in different stacking states.
The registry-dependent KC potential improves on this and correctly captures the energy variation,
but fails to yield reasonable forces. In particular, the KC model does not distinguish forces
on atoms in the AA and AB stacking states that are different in DFT calculations.
The KC model is also discontinuous at the cutoff, which can lead to difficulties
in energy minimization and loss of energy conservation in dynamic applications.

To address these limitations, we developed a new potential for graphitic structures based on the
KC model.
The \ipfullname (\ipshortname) has a smooth cutoff
and includes a dihedral-angle-dependent term to distinguish different stacking
states and obtain accurate forces.
The potential parameters were determined by training on a set of energies
and forces for a graphene bilayer at different layer spacing, sliding and
twisting, computed using GGA-DFT calculations, augmented with the MBD dispersion
correction to account for the long-range vdW interactions.

To test the quality of the potential,
we employed it to compute energetics, forces, and structural and elastic properties for a graphene bilayer in different states and graphite.
The validation tests show that compared with first-principles results:
\begin{enumerate}
\item \ipshortname correctly predicts the equilibrium layer spacing, interlayer binding energy, and generalized stacking fault energy of a graphene bilayer, as well as the equilibrium layer spacing of graphite.
\item \mc{\ipshortname underestimates the $c$-axis elastic modulus $C_{33}$ of graphite by about 10\%
relative to ACFDT-RPA and experiments, but this result
is in good agreement with PBE+MBD to which \ipshortname was fit.}
\item \ipshortname provides more accurate forces than the KC model across the entire range of bilayer rotations and in particular distinguishes the forces in the AA and AB states that the KC potential cannot.
\end{enumerate}

In two large-scale applications, not amenable to DFT calculations,
we showed that
\ipshortname and the KC potential agree qualitatively, but differ quantitatively
by 26\% in the out-of-plane relaxation of a twisted graphene bilayer,
and by 23\% in the normal force required to peel one graphene layer off graphite.

The added four-body dihedral-angle-dependent correction in \ipshortname
is very short-ranged ($\rho_\text{cut}=1.562$~\AA) and therefore the
computational overhead relative to KC is small. In fact, for the large-scale
applications (bilayer relaxation and peeling) described in \sref{sec:app},
\ipshortname was actually faster than the KC potential in terms of the
overall computation time due to improved convergence.

Although \ipshortname was parameterized against a training set consisting of graphene bilayers, it can be used to describe interlayer interactions for other systems such as graphite and multi-walled carbon nanotubes where the carbon atoms are arranged in layers.
This potential only provides a description of the interlayer interactions, and therefore must be used together with a companion model that provides the intralayer interactions, such as the Tersoff\cite{tersoff1988empirical,tersoff1989modeling} or REBO\cite{brenner1990physical,brenner2002rebo} potentials.
The \ipshortname functional form and associated carbon parameterization
are archived in the
OpenKIM repository\cite{drip_driver,drip_model,tadmor2011kim} at \url{https://openkim.org}.
They can be used with any KIM-compliant molecular simulation code,
see Appendix~\ref{appendix:kim} for details.

\begin{acknowledgments}
This research was partly supported by the Army Research Office (W911NF-14-1-0247) under the MURI program, and the National Science Foundation (NSF) under grants No.~DMR-1408211 and DMR-1408717.
The authors wish to acknowledge the Minnesota Supercomputing Institute (MSI) at the University of Minnesota for providing resources that contributed to the results reported in this paper.
M.~W.~thanks the University of Minnesota Doctoral Dissertation Fellowship for supporting his research.
The authors thank Alexey Kolmogorov for reading the manuscript and for his valuable comments.
\end{acknowledgments}

\appendix
\section{Using the Open Knowledgebase of Interatomic Models (OpenKIM)}
\label{appendix:kim}

The Open Knowledgebase of Interatomic Models (OpenKIM) (\url{https://openkim.org}) is an open-source, publicly accessible repository of classical interatomic potentials, as well as their predictions for material properties that can be visualized and compared with first-principles data. Interatomic potentials stored in OpenKIM that are compatible with the KIM application programming interface (API) are called ``KIM Models.'' KIM Models will work seamlessly with a variety of major simulation codes that are compatible with this standard including
LAMMPS\cite{plimpton1995fast,lammps},
ASE\cite{larsen2017atomic,ase},
DL\_POLY\cite{dlpoly}, and
GULP\cite{gale1997gulp,gulp}.

As an example, we describe how a KIM Model would be used with LAMMPS. In LAMMPS, reactive interatomic potentials are specified using the \verb|pair_style| command. LAMMPS has a ``\verb|pair_style kim|'' option for using KIM Models.  To use KIM Models with LAMMPS, perform the following steps:
\begin{enumerate}
\item Install the KIM API (see instructions at \url{https://openkim.org/kim-api/});
\item Download and install the desired potential from \url{https://openkim.org/} (see instructions that come with the API);
\item Enable KIM Models in LAMMPS by typing: ``\verb|make yes-kim|'' and then compiling LAMMPS.
\end{enumerate}
In a LAMMPS input script, a KIM Model is then selected in the same way as other LAMMPS potentials. For example, the potential developed in this paper can be used with the following two commands:
\begin{widetext}
  \quad\verb|pair_style  kim  DRIP_WenTadmor_2018_C__MO_070247075036_000|

  \quad\verb|pair_coeff  * *  C|
\end{widetext}
To use it together with another potential for the intralayer interactions, such as Tersoff\cite{tersoff1988empirical,tersoff1989modeling} or REBO\cite{brenner1990physical,brenner2002rebo}, use the LAMMPS ``\verb|pair_style hybrid/overlay|'' command (see the LAMMPS manual for details).

The advantage of releasing a potential as a KIM Model (as opposed to
just a file compatible with LAMMPS or another code), is that it will work with
not just LAMMPS, but other major codes as noted above.
In addition, a KIM Model has a ``KIM ID'' that can be cited in publications.
The KIM ID provides a unique permanent link to the archived content and includes a
three-digit version number to track changes. For example, a modification
to the model parameters would lead to a version upgrade (or a new forked model if appropriate).
Citing a KIM ID in a publication makes it possible for the reader to download the exact
potential used in the reported simulation and to reproduce the results.

%

\end{document}